\documentclass[apj]{emulateapj}
\include{epsf}
\usepackage{longtable}

\def\gsim{~\rlap{$>$}{\lower 1.0ex\hbox{$\sim$}}}
\def\lsim{~\rlap{$<$}{\lower 1.0ex\hbox{$\sim$}}} 
\def\Lsun{\hbox{$\rm\thinspace L_{\odot}$}}

\def\g{g_{\rm \thinspace 475}}
\def\z{z_{\rm \thinspace 850}}
\def\gz{g_{\rm \thinspace 475}-z_{\rm \thinspace 850}}

\def\deg{\mbox{$^{\circ}$}}

\shorttitle{Virgo Dwarf Globular Clusters}
\shortauthors{Beasley et al.}

\begin{document}


\title{The Globular Cluster System of the Virgo Dwarf Elliptical Galaxy VCC~1087\altaffilmark{1}}


\author{Michael A. Beasley, Jay Strader, Jean P. Brodie, A. Javier Cenarro}
\affil{UCO/Lick Observatory, University of California, Santa Cruz, CA 95064}
\email{mbeasley@ucolick.org, strader@ucolick.org, brodie@ucolick.org, cenarro@ucolick.org}




\and 

\author{M. Geha\altaffilmark{2}}
\affil{Carnegie Observatories, 813 Santa Barbara St., Pasadena, CA 91101}
\email{mgeha@ociw.edu}

\altaffiltext{1}{Some of the data presented herein were obtained at the W.M. Keck Observatory, which is 
operated as a scientific partnership among the California Institute of Technology, the 
University of California and the National Aeronautics and Space Administration. 
The Observatory was made possible by the generous financial support of the W.M. Keck Foundation.}
\altaffiltext{2}{Hubble Fellow}


\begin{abstract}
We present an analysis of the globular cluster (GC) system of the 
nucleated dwarf elliptical galaxy VCC~1087 in the Virgo cluster, based on 
Keck/LRIS spectroscopy and archival 
{\it Hubble Space Telescope}/Advanced Camera for Surveys imaging.
We estimate VCC~1087 hosts a total population of $77\pm19$ GCs, which 
corresponds to a relatively high $V$-band specific frequency of $5.8\pm1.4$.
The $\gz$ color distribution of the GCs shows 
a blue (metal-poor) peak with a tail of redder (metal-rich) clusters similar
in color to those seen in luminous ellipticals. 
The luminosity function of the GCs is log-normal and peaks 
at $M_{\g}^{\rm TO}=-7.2\pm0.3$, $M_{\z}^{\rm TO}=-8.1\pm0.2$.
These peak positions are consistent with those found for luminous Virgo ellipticals, 
suggesting either the lack of, or surprisingly similar, dynamical destruction
processes of GCs among dwarf and giant galaxies.
Spectroscopy of a subsample of 12 GCs suggests that the GC system
is old and coeval ($\gsim$10 Gyr), with a fairly broad metallicity
distribution (--1.8$\lsim$[m/H]$\lsim$--0.8). In contrast, an integrated
spectrum of the underlying galaxy starlight reveals that 
its optical luminosity is dominated by metal-rich, intermediate-aged stars.
Radial velocities of the GCs suggest rotation close 
to the major axis of the galaxy, and this rotation is
dynamically significant with $(v_{\rm rot}/\sigma_{\rm los})^*>1$.
A compilation of the kinematics of the GC systems of 9 early-type
galaxies shows surprising diversity in the $(v_{\rm rot}/\sigma_{\rm los})$
parameter for GC systems. In this context, the GC system of VCC~1087 exhibits 
the most significant rotation to velocity dispersion signature.
Dynamical mass modeling of the velocity dispersion profile
of the GCs and galaxy stars suggest fairly constant mass-to-light ratios of 
$\sim3$ out to 6.5 kpc. The present observations 
can entertain both baryonic and non-baryonic solutions, and 
GC velocities at larger radii would be most valuable with regard
to this issue.
Finally, we discuss the evolution of VCC~1087 in terms 
of the galaxy ``harassment'' scenario, and conclude that 
this galaxy may well be the remains of a faded, tidally
perturbed Sc spiral.
\end{abstract}



\section{Introduction}
\label{Introduction}

Characterized by their 
faint luminosities (M$_V\geq-18$) and low 
effective surface brightnesses ($\mu_{V,\rm eff}>22$), 
dwarf elliptical (dE) galaxies are numerically the dominant 
galaxy type in galaxy clusters 
(Binggeli, Sandage \& Tammann 1988).
dE galaxies are often subdivided into those
with nuclei (dE,N) and those without (dE,noN), a division
with a broad luminosity dependence in the sense
that most dEs brighter than M$_V\sim-16$ have
nuclei, while those fainter than M$_V\sim-12$
do not (Sandage, Binggeli \& Tamman 1985).
Recently it has come to light that
a further division may be made; one of
rotating and non-rotating dEs (De Rijcke et al.~2001; 
Pedraz et al.~2002; Geha et al.~2002, 2003).
Geha et al.~(2002, 2003) have shown that, 
so far, this distinction in dE galaxy kinematics
does not extend to their stellar populations 
or morphologies.

Formation scenarios for dEs 
must satisfy the observational constraints 
that (Conselice 2004): {\it (i)} dEs preferentially exist
in dense regions (even though they are
spatially diffuse), {\it (ii)} dEs exhibit both
rotational and pressure support, and  
{\it (iii)} dEs contain old, metal-poor stars 
and/or young, metal-rich stars.
The presently favored models of dE formation may be 
roughly distilled as follows: dEs are primordial 
objects which formed early on in dense environments 
(e.g., White \& Frenk 1991) and possibly ``squelched'' by 
reionisation (Tully et al.~2002), or that dEs are objects which had a progenitor population
(spiral or irregular galaxies) and were morphologically transformed
in the cluster environment (Mao \& Mo 1998; Moore et al.~1998; 
Mastropietro et al.~2005).
Observations of the velocity and spatial distributions
of dEs in galaxy clusters suggest that they may have 
infallen quite recently (Conselice et al.~2001). This 
combined with the fact that the stellar populations of dEs
differ from field dwarfs (van Zee et al.~2004a), 
and occasionally show spiral structures and \ion{H}{1}
(e.g., Jerjen et al.~2000; Barazza et al.~2002; Graham et al.~2003; De Rijcke et al.~2003)
supports the notion that at least some dEs were morphologically
transformed from late-type galaxies.

Globular clusters (GCs) provide a different perspective
with which to investigate dE galaxy formation. 
GCs are among the oldest observable stellar populations. 
Their high binding energies makes them relatively robust against tidal disruption, and 
their luminous, compact nature allows for easy separation from the 
galaxy background.
These and other factors make them ideal tracers of the mass aggregation history
of galaxies (e.g., West et al.~2004).
Durrell et al.~(1996a)
obtained deep $R$-band imaging of 11 Virgo dEs and 
found that the specific frequencies (S$_{\rm N}$) of
their sample ranged mostly from 3--8, similar to luminous elliptical 
galaxies, but considerably higher than the S$_{\rm N}\leq1$
of late-type disks. A {\it Hubble Space Telescope (HST)}/WFPC2 study by 
Miller et al.~(1998) (24 Virgo, Fornax \& Leo dEs) confirmed these findings, 
and found differences between the S$_{\rm N}$ of dE,Ns 
($\langle$ S$_{\rm N}\rangle=6.5\pm1.2$) and 
dE,noNs ($\langle$ S$_{\rm N}\rangle=3.1\pm0.5$).
Miller et al.~also found that, for dE,Ns, 
S$_{\rm N}$ increased with M$_V$, but little or no 
such correlation was seen in the dE,noN class.
The above results lead Miller et al.~(1998) to argue that 
the progenitors of dEs were probably a heterogeneous population, 
possibly reflected by the presence (or lack thereof) of dE nuclei.
Lotz et al.~(2004) examined the relationship between 
the colors of the GC systems and those of the galaxy envelope 
and nucleus using a larger sample of 69 Virgo, Fornax \& Leo dEs
imaged with {\it HST}/WFPC2. These authors found that the (photometric) 
metallicities of the dE GC systems correlated with host galaxy magnitude 
as $Z_{\rm GC}\propto L_B^{0.22}$, suggesting that the GCs ``knew''
about the host galaxy to which they would ultimately belong.
It now seems clear that, similar to the situation for ``red'' GCs
in luminous ellipticals, the mean colors of the ``blue'' GCs are correlated 
with host galaxy magnitude (Larsen et al.~2001; Strader et al.~2004, 2005; 
Peng et al.~2005).

In this paper we present a photometric and spectroscopic
analysis of the GC system of VCC~1087,  
a nucleated dwarf elliptical (dE3,N) located in projection 
$\sim$300 kpc southwest of the cD galaxy M87 (VCC: Virgo Cluster Catalog 
of Binggeli, Sandage \& Tamman 1985).
This paper is organized as follows:~in Section~\ref{Data} 
we discuss the observations
and reductions of the data used in this paper. 
Section~\ref{Analysis} presents
our analysis of the photometric and spectroscopic
data.  In Section~\ref{Discussion} 
we discuss the results of this study, and summarize
our findings in Section~\ref{Summary}. 

We note that Jerjen et al.~(2004) obtain two different distance estimates
for VCC~1087 from surface brightness fluctuations of 
$(m-M)_{\circ}=31.27\pm0.14$ and $31.39\pm0.18$, which
correspond to M$_V$=--17.8 and --18.0 respectively.
Unless otherwise stated, we adopted the shorter distance 
modulus which puts VCC~1087 some 2.7 Mpc beyond
the {\it HST} Key Project distance to Virgo
($(m-M)_{\circ}=30.92$; Freedman et al.~2001).

\section{The Data}
\label{Data}

\subsection{Imaging}
\label{Imaging}

Photometry of VCC~1087 and its GC system was derived 
from Sloan-type $g$-band (F475W) and $z$-band (F850LP) wide-field images taken 
with the Advanced Camera for Surveys (ACS) as part of the 
ACS Virgo Cluster Survey (C{\^ o}t{\' e} et al.~2004).
We follow the convention of C{\^ o}t{\' e} et al.~(2004)
and refer to the ACS F475W and F850LP filters as
$\g$ and $\z$ respectively.

Images were processed through the standard ACS pipeline.
\emph{Multidrizzle} was then used to combine individual images 
and reject cosmic rays.
Candidate GCs were selected as $4 \sigma$ detections
(matched in the $\g$ and $\z$ bands)
identified from $20\times20$ pixel median-subtracted images. 
Aperture photometry in a 5-pixel aperture was then obtained for all 
detected objects using the PHOT task in DAOPHOT II.  
Aperture corrections to a 10-pixel aperture were calculated 
from bright, isolated objects in VCC~1226, VCC~1316, 
VCC~1978, VCC~881 and VCC~798 (the five brightest 
galaxies in the ACS/Virgo survey of C{\^ o}t{\' e} et al.~2004). 
These magnitudes were then corrected to a nominal infinite aperture 
using the values of Sirianni et al.~(2005).  
We used photometric zeropoints 
from Sirianni et al.~to convert from the instrumental magnitude scale 
to the AB magnitude system, correcting for Galactic foreground extinction using 
the reddening curves of Cardelli et al.~(1989) and the 
DIRBE dust maps of Schlegel et al.~(1998). Photometric completeness limits
were determined from artificial star tests. The $50\%$ 
completeness limit in $\g$ is 26.6, and in $\z$ is 25.5. 

On the ACS images, GCs are round, compact objects; 
they can be separated from the majority of foreground stars 
and background galaxies using shape, color and size information.
GC candidates were selected as objects detected in both
$\g$ and $\z$ images to within 0.1 arcsec tolerance, with 
roundness and sharpness parameters (from DAOPHOT II)
in the intervals --0.5$\leq$~roundness~$\leq$0.5 and 
0.55$\leq$~sharpness~$\leq$0.90 respectively.
A color range of $0.5\leq(\gz)\leq2.0$ was adopted for GCs, 
which encompasses age and metallicity combinations of 
1 Gyr, [m/H]=--2.25 to 15 Gyr, [m/H]=0.67 (Maraston 1998).
Objects with $\g\leq19$ and $\z\leq18$ were
deemed too luminous to be true GCs. These bright limits
are roughly two magnitudes brighter than 
the most luminous Milky Way GC ($\omega$ Cen) at the
distance of VCC~1087.
Sizes for the GC candidates were measured using ISHAPE (Larsen 1999), 
and this information was used to expunge objects outside the 
range of half-light radius ($r_{\rm h}$) expected for 
Milky Way GCs ($0.5\leq r_{\rm h} {\rm (pc)} \leq 12$).
Finally, the GC sample was visually inspected for any obvious
non-GC properties (e.g., visible structures, blends) 
and these were culled from the final sample.

For $\g\leq24$, GCs are easily distinguished from 
galaxies since they appear more compact than galaxies for a given 
magnitude. Indeed, from our spectroscopy (see Section~\ref{Spectroscopy}), all 
the GCs identified on the {\it HST}/ACS images were found to 
be {\it bona fide} clusters (the faintest being $\g=23.7$).
Therefore, we conclude that the contamination
rate for cluster candidates brighter than this (approximately 
the turn-over (TO) in the GCLF) is consistent with zero. 
To quantify the contamination from background 
galaxies at magnitudes fainter than $\g=24$, 
a master background file was constructed from the ACS/Virgo survey
dwarf galaxy fields at large radii as discussed in Strader et al.~(2005).
From a region totalling 41 sqr. arcmin, we identify 37 objects 
which pass our criteria for GCs where we expect none (although note that, for the 
most luminous dEs such as VCC~1087, the GC systems do extend to the
edge of the ACS field as confirmed by our spectroscopy,  e.g.,  
Figure~\ref{Finder}). Scaling to one ACS pointing, we find
that $\sim10$ objects may be background galaxies.
We detect 68 GCs associated with VCC~1087. 
Based upon the incompleteness at the faint end 
of the GC luminosity function (GCLF) (Section~\ref{LF}), and correcting 
for incomplete areal coverage we estimate we have identified
$\sim88\%$ of the total GC population in this galaxy.
Therefore we estimate a total GC population of 77 GCs.
Our estimate of the contamination due to background galaxies
suggests that we may be overestimating the GC population
by $\sim13\%$. Rather than correcting for the total number of
GCs, we fold this absolute uncertainty into our population
estimate to yield $77\pm19$ GCs. This number excludes the nucleus of the galaxy (which meets
all of our GC selection criteria.)
Adopting M$_V$=--17.8 for VCC~1087 
(Jerjen et al.~2004),  we obtain a specific frequency of $5.8\pm1.4$. 

\subsection{Spectroscopy}
\label{Spectroscopy}

Optical spectra were obtained for 12 GC candidates
identified from the $\g$ and $\z$ {\it HST}/ACS images (Section~\ref{Imaging})
during the nights of 24-25 April, 2004, using the Low Resolution 
Imaging Spectrograph (LRIS) (Oke et al.~1995) on the Keck I telescope.
A total of 15 1800s exposures were obtained 
(totaling 7.5 hours) through a 0.8 arcsec slitmask. 
We also obtained a 300s, 0.8 arcsec long-slit spectrum of 
VCC~1087 oriented along the major axis
of the galaxy (P.A.=104$\deg$).
A flux standard (BD+33$\deg$2642) taken from
Oke (1990) was observed with a longslit to allow
first-order continuum corrections, along with a
number of radial velocity and Lick/IDS standard stars (from
Worthey et al.~1994). Bias frames, dome/sky flats 
and Cadmium-Neon-Argon  arcs were taken
at various points during the run for calibration purposes.
The seeing varied from 0.6-0.9 arcsec during the two-night
run.

We obtained simultaneous blue and red spectra through the
use of a dichroic which split the beam at 6800\AA.
On the blue side, a 600 lines/mm grism blazed at 4000\AA\ was used 
to yield an effective wavelength range of 3300--5900\AA\ 
and a resolution of $\sim$3\AA\ (FWHM).
This effective wavelength range varied slightly between objects
due to the varying slitlet positions on the slitmask.
On the red side, a 831 lines/mm grating was used, 
blazed at 8200\AA\ to cover the Ca II triplet region.
The analysis of the Ca II triplet data will be discussed 
elsewhere.

The reduction of these data was performed with 
IRAF\footnote{IRAF is distributed by the National Optical Astronomy 
Observatory, which is operated by the Association of Universities for 
Research in Astronomy, Inc., under cooperative agreement with the 
National Science Foundation.}.
All science images were bias-subtracted and then flat-fielded with 
twilight sky flats which were co-added and normalized.
Spectra were optimally extracted (see Horne 1986) and wavelength 
calibrated with solutions obtained from the arc exposures. 
Wavelength residuals of 0.2~\AA\ were typical. The zeropoints
of the wavelength calibration were checked against skylines in the 
background spectra, and were in some cases adjusted by a few tenths of 
an Angstrom. These wavelength solutions were further adjusted as described below.
The spectra were then combined with cosmic-ray rejection, and flux
calibrated using the response function derived from our flux standard.
Error spectra were extracted in parallel with the object
spectra to facilitate accurate estimation of 
uncertainties in our analysis.

The velocity dispersion of the VCC~1087 GC system
was anticipated to be similar to that of the galaxy 
itself at $\sim40$ km s$^{-1}$ (Geha et al.~2003). This required that 
particular care was taken to characterize the uncertainties
in our radial velocity measurements in this study.
The theoretical velocity accuracy for 
our spectral resolution is $\sim20$ km s$^{-1}$
at 4500\AA. Radial velocities for the GCs were derived using the 
MOVEL algorithm described in Gonz\'alez~(1993).
This is an improvement over the classic
Fourier quotient method (Sargent et al.~1977) which determines radial velocities
and velocity dispersions simultaneously.
Although our spectral resolution did not allow to us measure velocity 
dispersions for the GCs themselves, we used this approach 
since experimentation showed that it provides more robust velocity determinations than
classical cross-correlation procedures. As reference template spectra
for this method, we employed a number of Lick stars observed during the
run. To prevent overestimating the uncertainties in the method, which
can appear when velocity dispersion solutions are close to zero, the
GC spectra were broadened by convolving them with a
Gaussian of $\sigma=50$ km s$^{-1}$. Uncertainties in the radial velocities 
were computed from 100 Montecarlo simulations using the error spectra. 

After measuring velocities, small
deviations from the linear wavelength scale were corrected by using
the MILES spectral energy distributions (Vazdekis et al.~2005, in preparation) as
rest-frame wavelength calibrators. Each GC spectrum was
cross-correlated with an appropriate MILES template in $\sim$100
small, overlapping spectral regions in the dispersion direction. The derived
offsets (around zero but not necessarily null) were fitted as a
function of wavelength, thus providing a low-order, polynomial
correction to the prior wavelength calibration that was applied to the
deredshifted data. In this way, all the final GC spectra resulted in having
an homogeneous, rest-frame, linear wavelength scale.
These velocities are listed in Table~\ref{tab1}.
From Table~\ref{tab1}, the mean velocity uncertainty is 22 km s$^{-1}$, 
in good agreement with the expected value.
Example spectra are shown in Figure~\ref{Spectra}.

Lick indices were measured from the spectra using the 
definitions given in Trager et al.~(2000) and 
Worthey \& Ottaviani (1997). Index uncertainties were determined 
from the error spectra.  Small additive offsets where applied to the 
measured indices obtained from our 
Lick standard star observations, and were in general smaller than
the index uncertainties.
 
\vbox{
\begin{center}
\leavevmode
\hbox{%
\epsfxsize=8cm
\epsffile{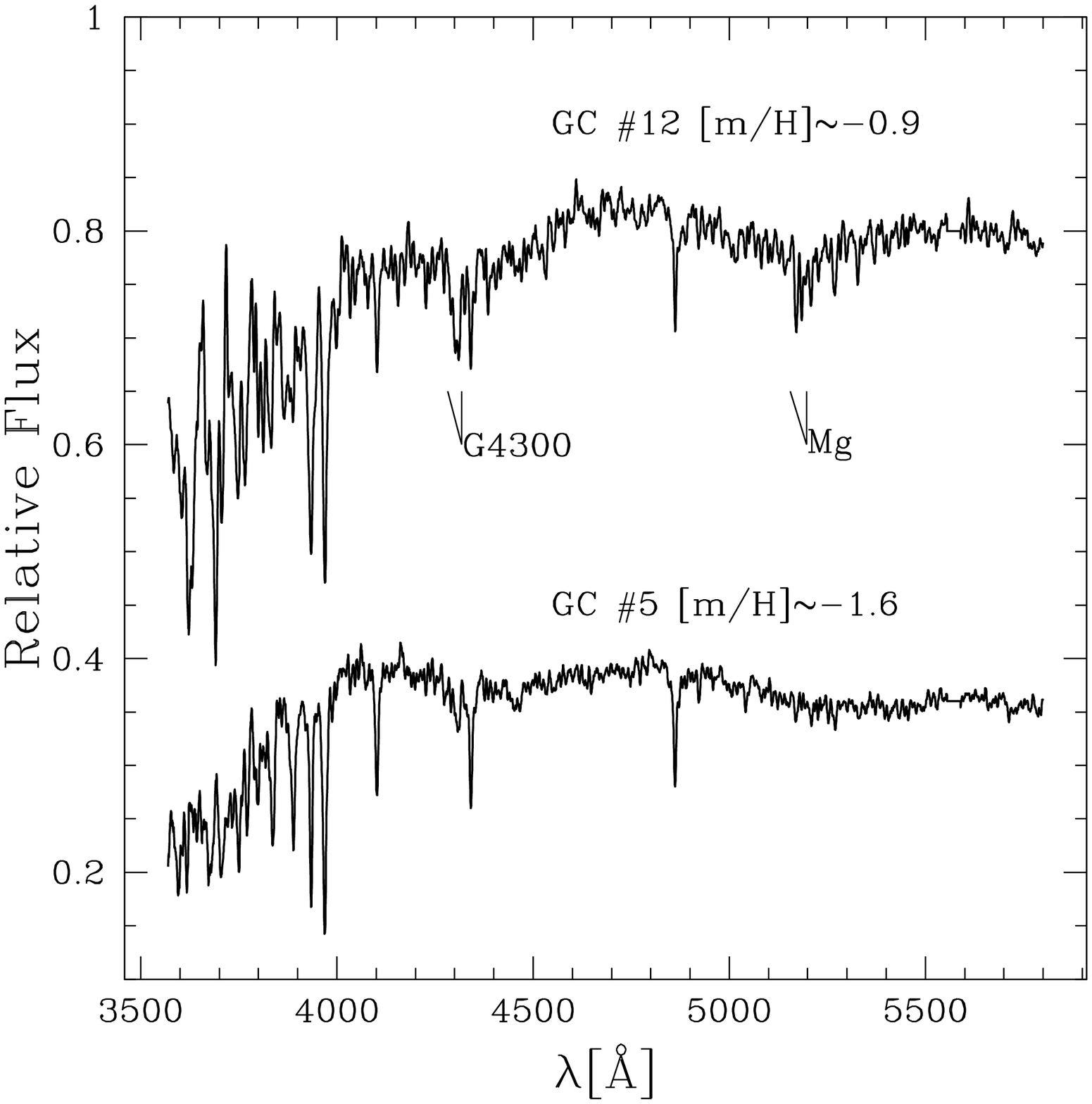}}
\figcaption{\small
Comparison of two VCC~1087 GC spectra one of intermediate metallicity (GC\#12; upper),
and one metal-poor (GC\#5; lower).  Magnesium hydride ($\sim$5170\AA) and
the G-band ($\sim$4300\AA) are stronger in the more metal-rich GC. 
\label{Spectra}}
\end{center}}

\section{Analysis}
\label{Analysis}

\subsection{Photometric Properties of the GC System}
\label{Photometry}

The spatial distribution of the GC candidates 
on the $\g$ image is shown in 
Figure~\ref{Finder}. The spectroscopically confirmed
GCs are also indicated (Table~\ref{tab1}).
The GCs are concentrated towards the center of the galaxy, and 
appear spatially elongated in the NE--SW sense.
This general impression is supported by the azimuthally-averaged
distribution of the GCs (Figure~\ref{Azimuth}).
A sinusoidal fit to the unbinned data gives a position angle (P.A., measured E through N) 
of $85\pm49\deg$.
Unfortunately the small sample size does not allow for stronger constraints
on the major axis P.A. of the GC system, 
but the P.A.'s of the GCs and the galaxy isophotes (104$\deg$) are consistent
with each other.
The major- to minor-axis ratio of the GC spatial
distribution gives an approximate ellipticity of $\epsilon\sim0.3$, which is
similar the galaxy ellipticity ($\epsilon=0.26$) beyond 1 kpc.

\vbox{
\begin{center}
\leavevmode
\hbox{%
\epsfxsize=10cm
\epsffile{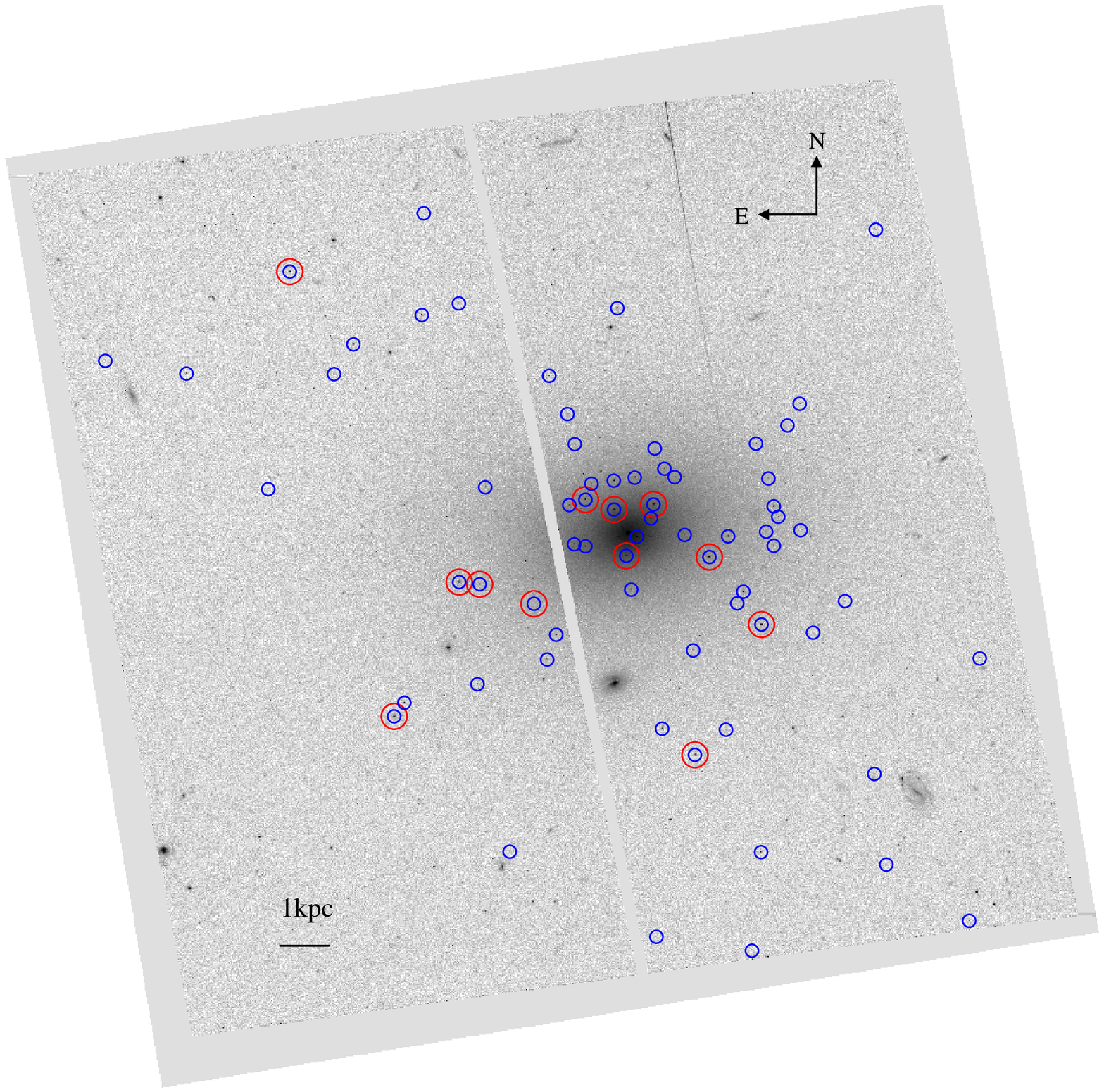}}
\figcaption{\small
{\it HST}/ACS $\g$ image of VCC~1087. GC candidates identified from our {\it HST}
photometry are indicated by small circles, those for which we have spectra and are 
positively identified as GCs are double-circled. The bright nucleus
of the galaxy can be seen in the center of VCC~1087, adjacent to the most central GC.
\label{Finder}}
\end{center}}

\vbox{
\begin{center}
\leavevmode
\hbox{%
\epsfxsize=8cm
\epsffile{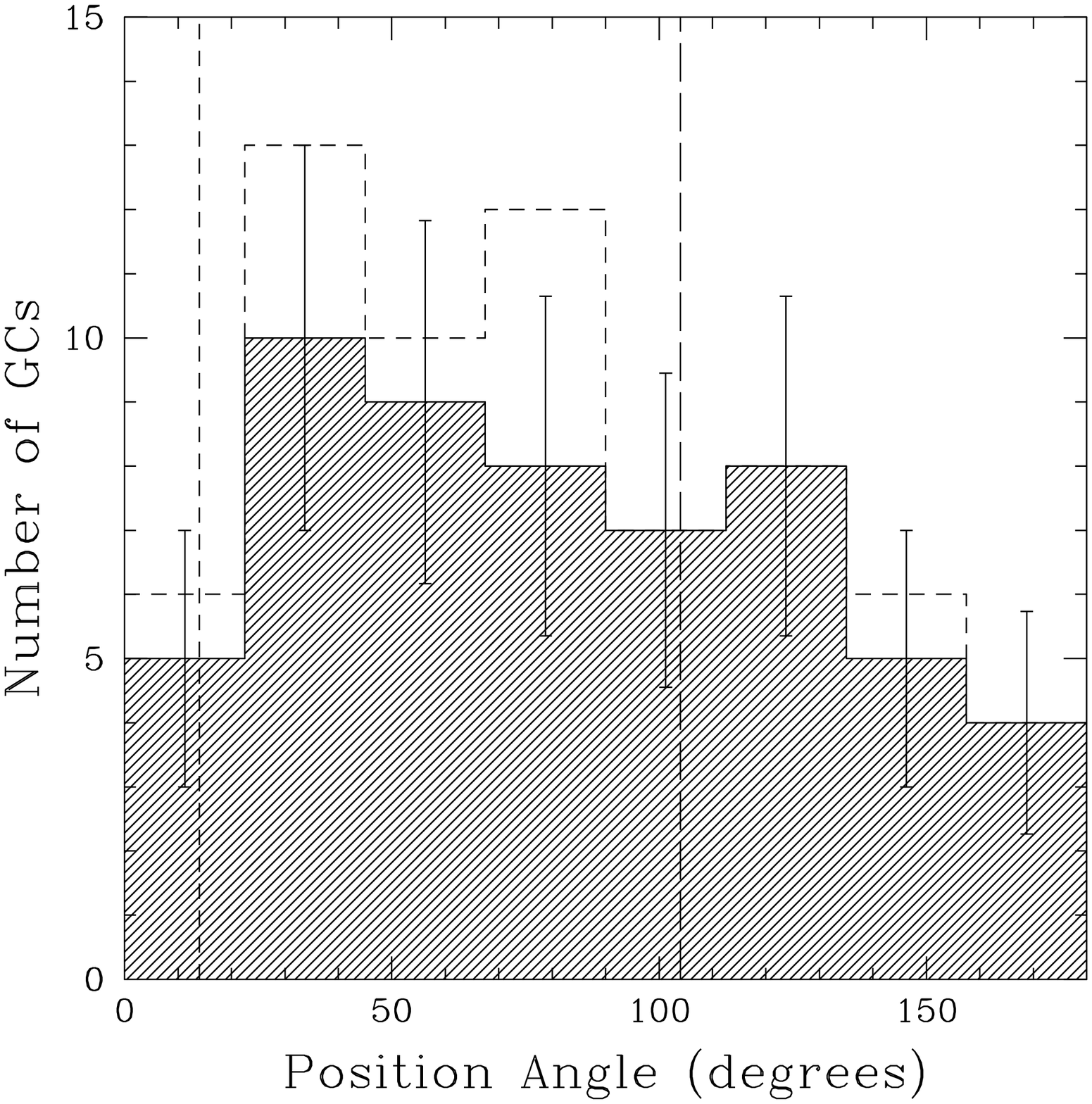}}
\figcaption{\small
Angular distributions of GCs associated with VCC~1087 
interior to 1.4 arcmin (the radius from the galaxy center
which permits full azimuthal coverage; shaded histogram) and for the
full sample of GCs (dashed line). 
The vertical dashed line represents the position angle of the major
axis of VCC~1087 (P.A.=104$\deg$), the short dashed
line indicates the position of the minor axis.
\label{Azimuth}}
\end{center}}

The nucleus of the galaxy is well resolved, and we measure
$\g=20.22\pm0.01$, $\gz=1.30\pm0.01$ (Table~\ref{tab1}).
We have also measured the envelope colors of the galaxy, 
and obtain $\gz=1.38\pm0.01$, indicating that the envelope
of the galaxy is marginally redder than its nucleus\footnote{
It was unnecessary to mask out the nucleus in measuring the 
envelope colors since it contributes only $\sim1\%$ to the 
total integrated light of the galaxy.}.
This is in agreement with the {\it HST} studies of De Propris et al.~(2005) and 
 Lotz et al.~(2004), but is in the opposite sense to the conclusions of 
ground-based work (e.g., Rakos \& Schombert 2004; 
Caldwell \& Bothun 1987). Presumably this disagreement stems
from the fact that the ground-based work failed to resolve the dE nuclei.

\vbox{
\begin{center}
\leavevmode
\hbox{%
\epsfxsize=8cm
\epsffile{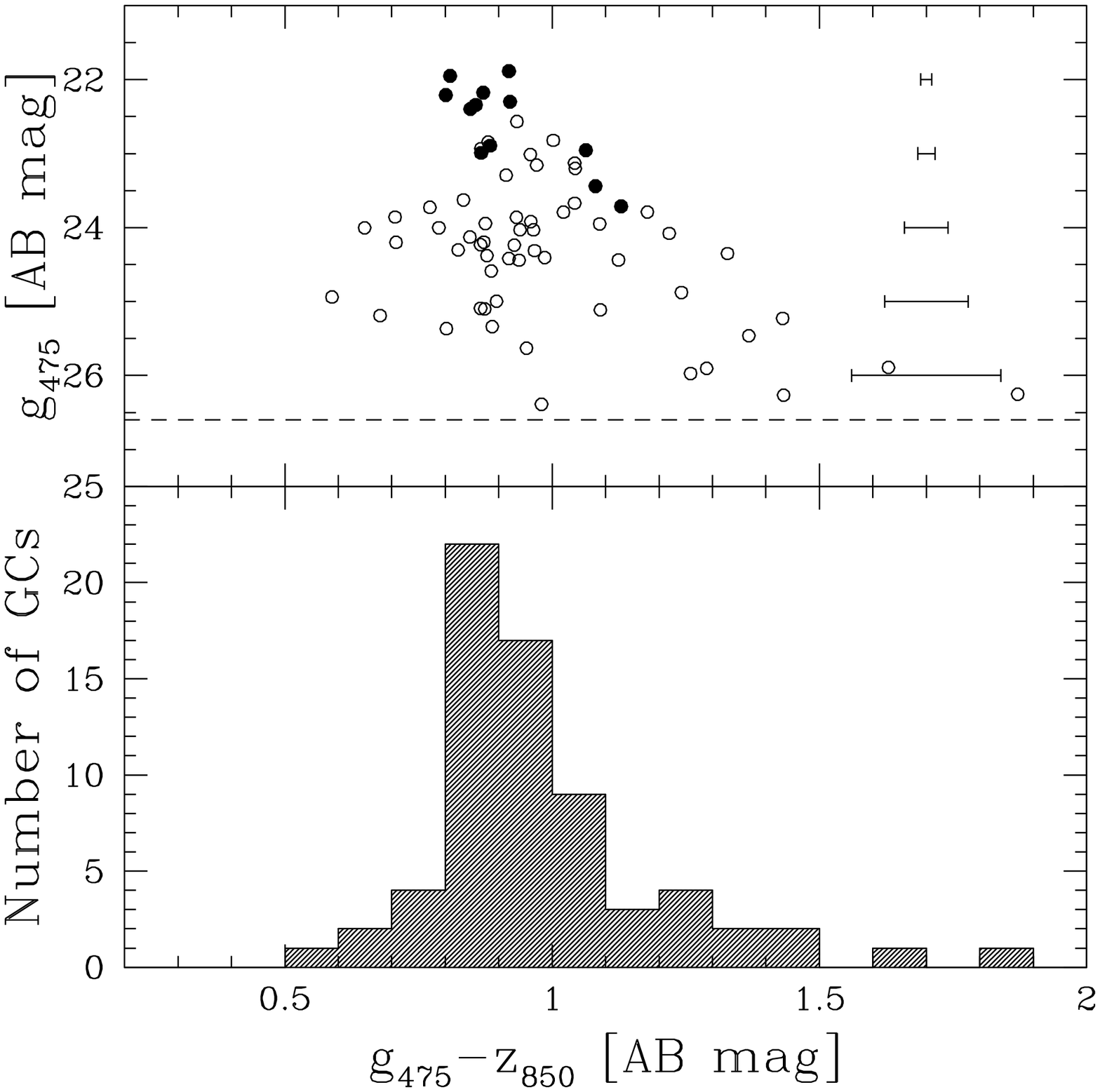}}
\figcaption{\small
Color-magnitude diagram for all GC candidates from {\it HST}/ACS images 
(open circles)
and those for which we have obtained spectra (filled circles). The 
horizontal dashed line (upper panel) shows the 50\% completeness limits of these
data. 
\label{CMD}}
\end{center}}

The color-magnitude distribution for the confirmed and 
candidate GCs is shown in Figure~\ref{CMD}. 
The GCs show a blue
peak (metal-poor clusters, henceforth MPCs), and a possible tail of red 
(metal-rich clusters, henceforth MRCs) GCs. 
A straight mean 
of the color distribution gives $\gz=0.98$ 
with a dispersion of 0.22 mag. 
Peak locations for the MPCs and (putative) MRCs were determined
using the Bayesian code Nmix (Richardson \& Green 1997), which 
implements mixture modeling of heteroscedastic normal distributions
with the number of subpopulations a free parameter.
For VCC~1087, the posterior distribution of this parameter 
peaked at two, suggesting two groups in the color distribution
with peaks at $\gz=0.90$ and 1.17 for the MPCs and MRCs
respectively. 

The red objects ($\gz\sim1.5$) near to our 50\% completeness
limit are possibly background galaxies (this statement can only
be confirmed spectroscopically). To check whether these
objects are driving this bimodal signature, we cut the 
CMD at $\g=25$ and re-ran the code. Again, two populations were
preferred with peaks at $\gz=0.90$ and 1.16. Similarly, a KMM test
(Ashman, Bird \& Zepf 1994) returned two peaks at 
$\gz=0.89$ and 1.16, although with a fairly low significance
($p=1.8$). These peak positions are consistent with the galaxy luminosity--
GC color relations derived by Strader et al.~(2005).
We tentatively identify the objects with $\g\sim24$ and 
$\gz\sim1.1$ as MRCs, three of which, confirmed by our spectroscopy,
are consistent with this interpretation (e.g., Figure~\ref{MH}).
Note that, in the subsequent analysis, we have not separated out the 
blue and red subpopulations due to the small
number of clusters.

The radial density profile of the GC system is compared to the 
galaxy surface brightness profile in Figure~\ref{profile}. We 
fitted the GC profile with the functional form $\rho \propto r^{-\alpha}$.
Excluding the innermost bin (see below), we find $\alpha=-1.1\pm0.4$, 
where the uncertainty comes from a bootstrap of the density profile.
A power-law index of $\alpha=-1.1\pm0.4$ is relatively shallow, 
more characteristic of the GC systems of cD galaxies than those of low-luminosity
ellipticals (e.g., Harris et al.~2004). Similarly shallow
density profiles have been suggested for the GC systems of dEs
previously (Durrell et al 1996a; Lotz et al.~2001).
The galaxy starlight is well fit by a S$\acute{e}$rsic (1968) 
profile with $n=1.4$ (Geha et al.~2003), where $n=1$ 
corresponds to an exponential
profile, and $n=4$ corresponds to an $r^{1/4}$ law.

There is the suggestion of a ``core'' in the GC density
profile near the galaxy effective radius, 
where there is a deficit of clusters compared to the power-law
slope. 
Lotz et al.~(2001) discussed the existence of such 
cores in the context of the formation of dE nuclei
through the dynamical destruction of GCs at small radii.
Could these ``missing'' GCs be the cause of the break in the 
density profile? For GCs at the 
turn-over of the GCLF (see below), $L_{\rm nucleus}/L_{\rm GC}\sim40$, 
i.e., 40 typical GCs would need to coalesce to give the 
observed nuclear luminosity. These additional GCs (added to the innermost 
radial bin) are indicated by the open square in Figure~\ref{profile} 
(for simplicity, we assume that these ``missing'' GCs occupied the same radial 
distribution as those present in the inner bin.)
The position of these GCs lie on an extrapolation 
of the power-law fit in Figure~\ref{profile}.
This would seem to suggest that the existence of a core may indicate 
dynamical destruction processes.

However, since the timescale for dynamical friction is expected to scale
as $M_{\rm GC}^{-1}$, one would expect that the most massive 
GCs decay first (for a given spatial density profile).
Drawing clusters from the bright end of the observed
GCLF (Figure~\ref{LF}), $\sim10$ GCs are required to equal
the luminosity of the galaxy nucleus 
(the number of luminous GCs required is clearly dependent upon 
the exact shape of the ``zero-age'' luminosity function, 
but presumably the bright end has not become brighter
with time). This is indicated by the open circle in
Figure~\ref{profile}, whose position is $\sim4~\sigma$
away from the extrapolated profile fit. 
From these very simplistic arguments, 
unless somehow turn-over mass clusters are preferentially
disrupted, the luminosity
of the nucleus and the core in the inner parts 
of GC surface density of VCC~1087 cannot both be explained 
purely due to the action of dynamical friction.

\vbox{
\begin{center}
\leavevmode
\hbox{%
\epsfxsize=8cm
\epsffile{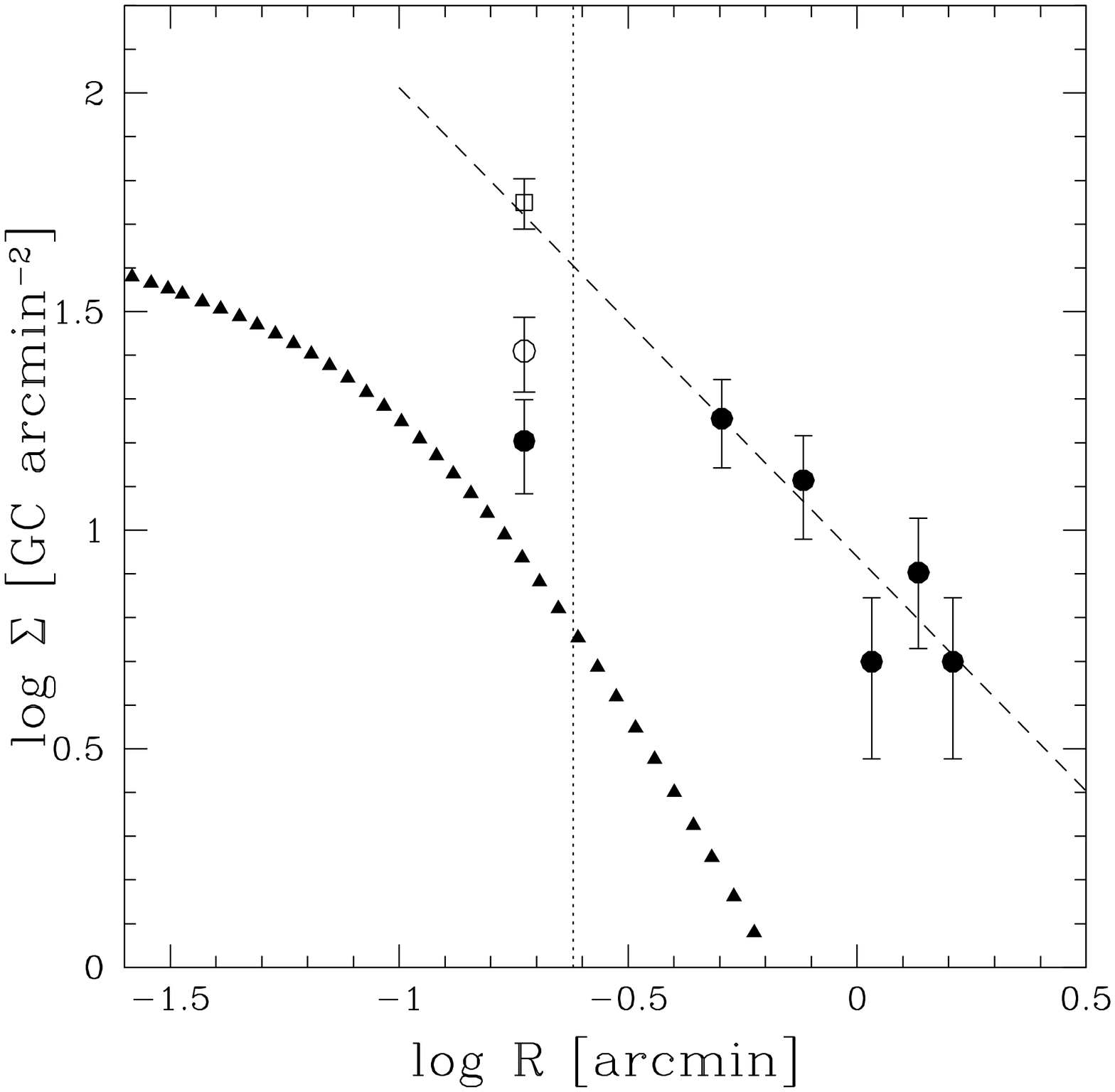}}
\figcaption{\small
Radial density profile of the VCC~1087 GC system (filled circles with 
Poisson error bars) compared
to the surface brightness profile of the galaxy
(arbitrarily normalized). The short dashed lines indicates 
a power-law fit to the GC data with exponent $\alpha=-1.1$, excluding 
the innermost bin.
The vertical dotted line indicates the effective radius of the 
galaxy at $\sim$1.2 kpc. The open square shows the 
expected surface density of GCs ``corrected'' for the effects
of dynamical friction for GCs at the turnover of the GCLF.
The open circle shows the same but for GCs at the bright end of the 
GCLF (see text).
\label{profile}}
\end{center}}

\vbox{
\begin{center}
\leavevmode
\hbox{%
\epsfxsize=8cm
\epsffile{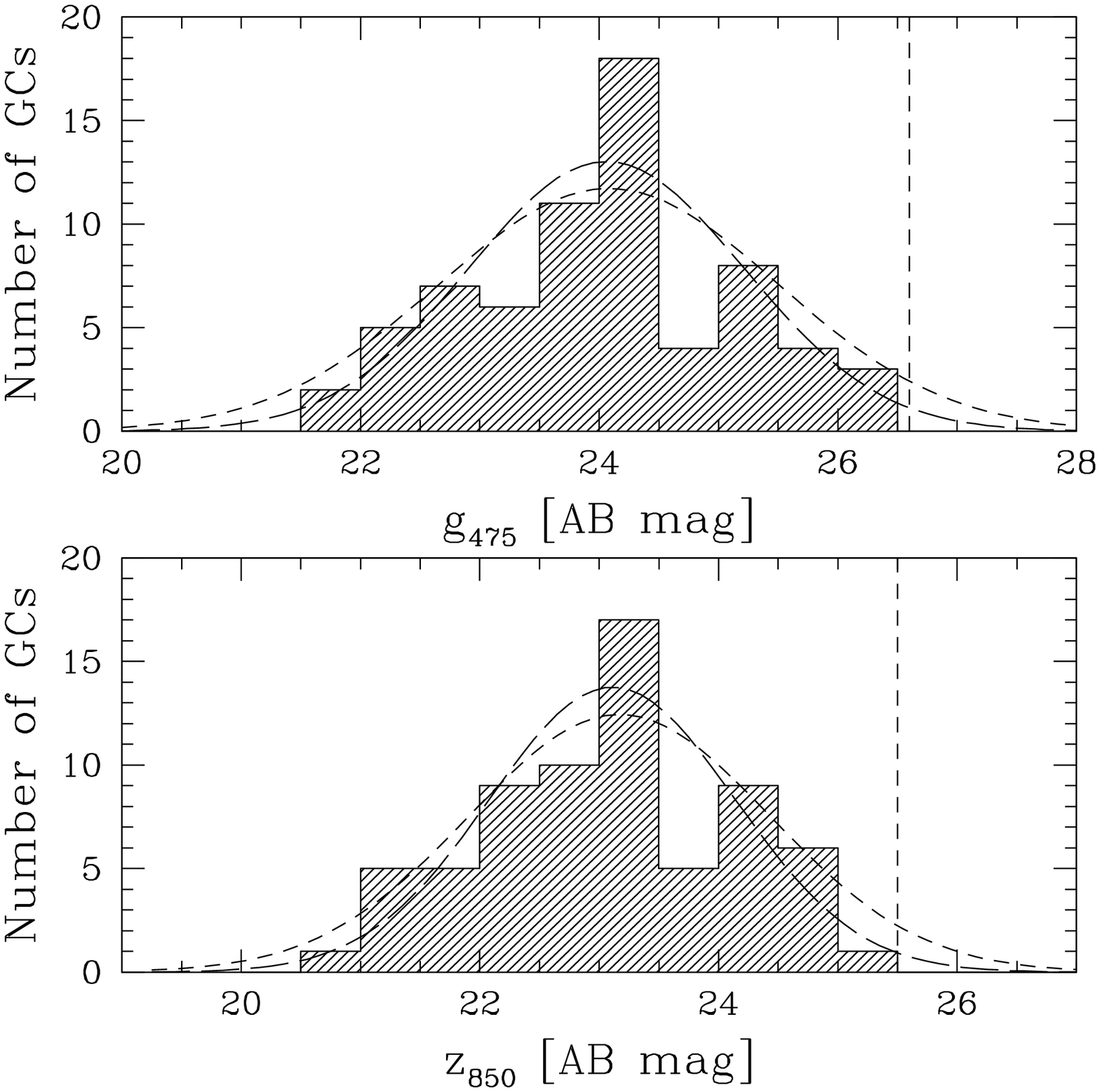}}
\figcaption{\small
Luminosity functions for the VCC~1087 GC system in $\g$ (top panel) and $\z$. 
Curves are maximum-likelihood estimates for the peaks and 
dispersions of the GCLFs assuming Gaussian distributions 
(short-dashed curves : the peak and dispersion are free parameters; 
long-dashed curves : the dispersions are fixed at 
$\sigma_{\g}=1.1$ and $\sigma_{\z}=1.0$).
The vertical dashed lines represent the 50\% completeness limit of the photometry.
\label{LF}}
\end{center}}

The luminosity functions for the VCC~1087 GCs 
in the $\g$ and $\z$ bands are shown in Figure~\ref{LF}.
The 50\% completeness limits of the photometry suggest
that we detect GCs down to the faint limit of
the GCLF.
We fitted both Gaussian and t$_5$ functions  
to the unbinned magnitude data using a slightly
modified version of the maximum-likelihood
code of Secker \& Harris (1993).
Within the uncertainties of the fits, 
both functions yield indistinguishable  means and dispersions
(assuming $\sigma_{t5}=0.775\sigma_{g}$), and we  
give the more familiar Gaussian best-fit parameters.
Leaving the Gaussian means and dispersions as free parameters
in the fit, we obtain $\g^{\rm TO}=24.1\pm0.3$, $\sigma_{\g}=1.4\pm0.3$
and $\z^{\rm TO}=23.2\pm0.2$, $\sigma_{\z}=1.26\pm0.3$.

These fits are shown in Figure~\ref{LF}. Inspection of the
figure suggests that the dispersions of the GCLF, particularly
in the case of $\g$, may be overestimated. 
Experimentation with the  Secker \& Harris (1993)
code suggested that for small $N$, $\sigma$ can become
artificially large. The straight standard deviations of the
$\g$ and $\z$ GCLFs (not accounting for incompleteness)
are 1.1 and 1.0 respectively, and we fixed these
in the code to assess the impact on the derived GCLF peak 
positions. $\g^{\rm TO}$ becomes 0.06 mag brighter, 
while $\z^{\rm TO}$ becomes 0.15 mag brighter. 
Both these shifts are within the uncertainties, and therefore
we adopt the original maximum-likelihood parameters.

At our adopted distance to VCC~1087, the 
apparent turnover magnitudes we derive correspond
to absolute magnitudes of $M_{\g}^{\rm TO}=-7.2\pm0.3$
and $M_{\z}^{\rm TO}=-8.1\pm0.2$.
These turn-over magnitudes are consistent with that found 
by Strader et al.~(2005), who constructed a
GCLF from 37 individual Virgo dEs. These authors
found $M_{\g}^{\rm TO}=-7.21\pm0.14$
and $M_{\z}^{\rm TO}=-8.14\pm0.15$, in which 
our photometry for VCC~1087 was included.

Interestingly, from the mean GCLF turnover magnitudes
of three giant Virgo ellipticals (NGC~4486, NGC~4472 \& NGC~4649),
Strader et al.~(2005) found $M_{\g}^{\rm TO}=-7.07\pm0.15$
and $M_{\z}^{\rm TO}=-8.19\pm0.15$, i.e., the Virgo
dEs (including VCC~1087) have turnover-magnitudes
consistent with luminous Virgo ellipticals.
This consistency, particularly in $\z$ for which color
differences in the systems are expected to have little
effect on the GCLF peak positions, seems to suggest 
that either dynamical destruction of GCs in dEs and luminous Es operates 
very similarly, or that it is not a dominant process in
shaping the mass function of GC systems. 
Dynamical friction (Chandrasekhar 1943) and 
disk/bulge shocking (Fall \& Rees 1977) are expected to operate
on the high- and low-mass ends of the GC mass function respectively.
These dynamical destruction processes are expected to be
functions of the dynamical structure and gravitational potential 
of the host galaxy (for a given GC mass function and spatial distribution), 
which makes it unlikely that dwarf and giant galaxies
will destroy their GCs similarly. Processes which may act 
against dynamical friction are discussed in Lotz et al. (2001), 
and include high dark matter halo densities and/or larger 
halo core radii with respect to the baryons, and kinetic heating 
mechanisms such as that which may come from GC scattering from black holes
or through tidal stirring. The affects of disk/bulge shocking could be 
ameliorated by placing the GCs on primarily tangential orbits, 
thereby avoiding the galaxy centers. These possibilities would be
a fruitful area of exploration through numerical simulation.

\subsection{Kinematics}
\label{Kinematics}

We obtain a biweight location (Beers et al.~1990) velocity for the GC system
of $686\pm24$ km s$^{-1}$. This compares well with the 
canonical NED value (Simien \& Prugniel 2002) of $675\pm12$ km s$^{-1}$, 
and with the velocity derived from our integrated spectrum
of the galaxy ($646\pm30$ km s$^{-1}$).
We have looked for rotation in the GC system by 
performing linear weighted fits to the GC velocities
as a function of position angle.
The GCs show a maximal rotation
signature about P.A.=127$\pm18\deg$, with $87\pm29$ km s$^{-1}$ arcmin$^{-1}$.
Over the radial extent of these data (i.e., measured
from the center of the galaxy to the outermost GC) this corresponds to 
V$_{\phi}=104\pm35$ km s$^{-1}$, assuming solid-body rotation.
The linear fit to the GC velocities
about P.A.$=127\deg$ is shown in Figure~\ref{rotation}.
The P.A. of the rotation axis of the GCs appears to be close to 
{\it parallel} to the P.A. of the galaxy isophotes ($104\deg$)
and the P.A. of the GCs ($85\pm49\deg$).
This alignment of the rotation axis is in a similar sense to that found by 
Puzia et al.~(2000) for the GC system of the dE NGC~3115 DW1. 
The GCs in VCC~1087 appear to rotate {\it along} an axis of $37\pm18\deg$, which is 
close to the minor axis P.A. of VCC~1087's isophotes ($\sim14\deg$).

\vbox{
\begin{center}
\leavevmode
\hbox{%
\epsfxsize=8cm
\epsffile{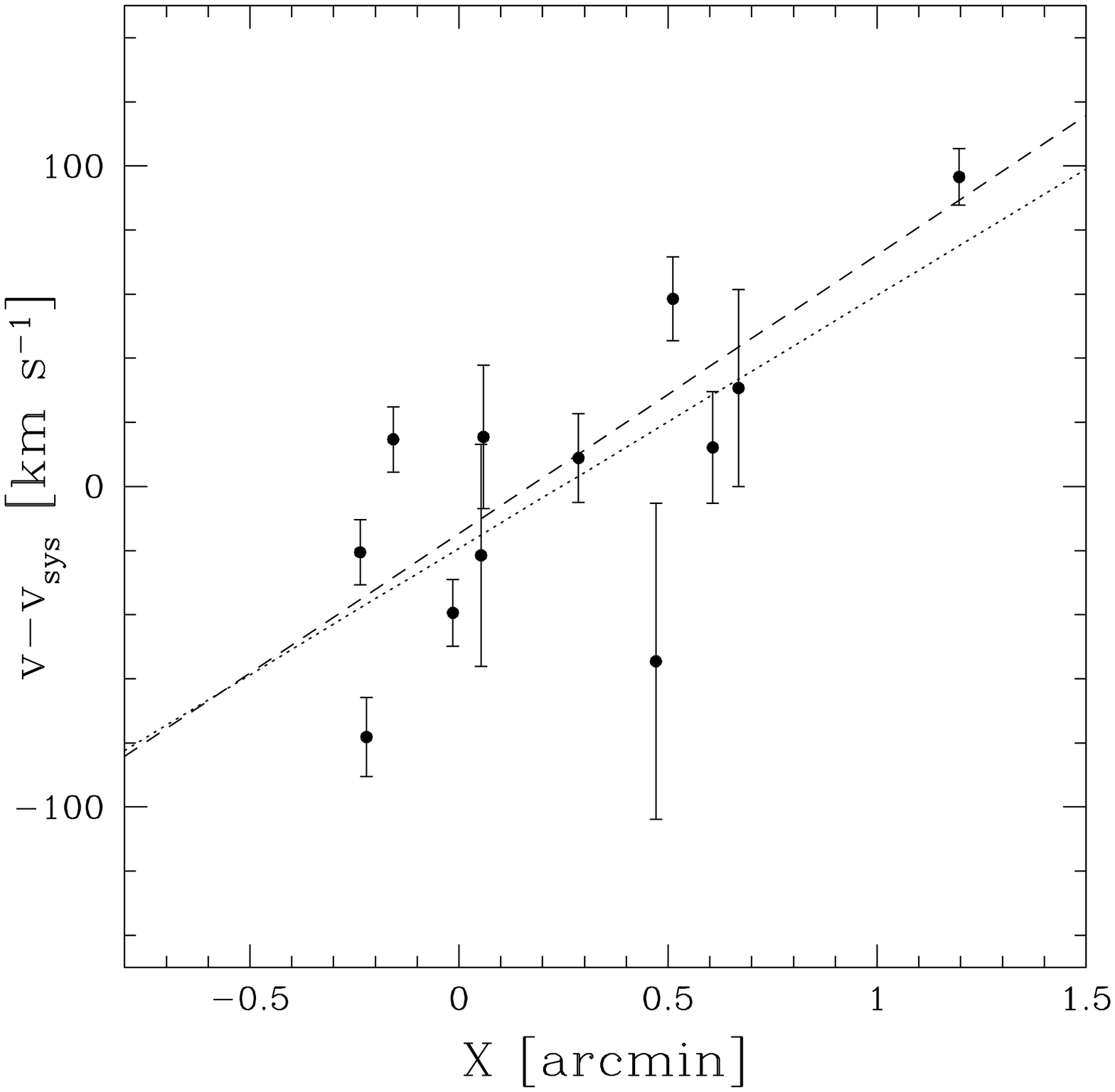}}
\figcaption{\small
Radial velocities of the GCs (corrected for a systemic velocity of
686 km s$^{-1}$) as a function of projected distance along a P.A. 
of 127$\deg$. The dashed line represents a weighted
linear fit to these data, suggesting rotation in the GC system of 
52$\pm$18 km s$^{-1}$. The dotted line indicates an unweighted
linear fit to these data and yields comparable rotation.
\label{rotation}}
\end{center}}

The biweight scale of the velocities (velocity dispersion, 
$\sigma_{\rm los}$) of the VCC~1087 GCs is $49\pm15$ km s$^{-1}$. 
To obtain a ``true''velocity dispersion  for the GCs 
(denoted $\sigma'_{\rm los}$) 
we corrected for our intrinsic velocity 
uncertainties (subtracting in quadrature) and for rotation in the GC system.
This yields $\sigma'_{\rm los}=29\pm6$ km s$^{-1}$ at a mean galactocentric 
radius of 35 arcsec ($\sim3$ kpc).
Similar to such considerations for elliptical
galaxies, one may ask whether the rotation in the GC system is 
dynamically significant by considering an isotropic
oblate rotator with ellipticity $\epsilon$ (e.g., Binney 1978):

\begin{equation}\label{eq:rot}
(v_{\phi}/\sigma_{\rm los})^*=\frac{(v_{\phi}/\sigma_{\rm los})_{\rm observed}}{(v_{\phi}/\sigma_{\rm los})_{\rm model}} = \frac{(v_{\phi}/\sigma_{\rm los})_{\rm observed}}{[\epsilon/(1-\epsilon)]^{1/2}}
\end{equation}

\noindent 
For a system with $\epsilon=0.3$ (approximately the ellipticity of both the galaxy 
 the GC system), $(v_{\phi}/\sigma_{\rm los})_{\rm model}=0.65$.
For the VCC~1087 GCs we find: $(v_{\phi}/\sigma^{'}_{\rm los})_{\rm observed}=3.6\pm1.8$, 
giving $(v_{\phi}/\sigma_{\rm los})^*=5.5\pm2.6$.
This result is suggestive that the GC system is  
supported by rotation.

We have investigated where the VCC~1087 GC system
lies with respect to other GC systems in the 
ellipticity--$(v_{\phi}/\sigma_{\rm los})$ plane.
There is surprisingly little information on the azimuthal
distributions GC systems in the literature, precluding 
an accurate knowledge of $\epsilon$.
We have assumed that the ellipticity of the
host galaxy reflects that of the GC system (and, where
they have been separated, that of the both the MPC and MRC 
subpopulations). There is some observational support 
that this assumption holds for the MRCs
(Dirsch et al.~2005; Lee et al.~1998) although for the MPCs
this is questionable (Kundu \& Whitmore 1998, but see 
Section~\ref{Photometry}).
In the Milky Way GC system, 
the (ensemble) halo GCs are broadly spherical, 
and the metal-rich GCs show flattening about the 
Galactic plane (e.g., Borkova \& Marsakov 2000).
These considerations notwithstanding, we plot 
the $(v_{\phi}/\sigma_{\rm los})$ ratios for
9 GC systems in Figure~\ref{vsig} using the 
data and sources given in Table~\ref{kin}.
 
Only the GC systems for early-type
galaxies are shown in Figure~\ref{vsig}, since 
$v_{\phi}/\sigma_{\rm los}$ is less meaningful
for disk systems (note that we classify NGC~4594--the Sombrero--as an S0 here.)
The GC systems
of the four E galaxies lie on or below the curve for rotationally
flattened systems (i.e., velocity dispersion dominates rotation), 
whereas the positions of the two dE 
galaxies suggests that rotation becomes increasingly 
important for these systems.
The kinematics of the GC systems which have been separated into MPC
and MRC subpopulations show significant diversity, with neither
subpopulation biased to either rotational or pressure support.
In all cases, if the MPCs are spherically distributed,
similar to the case for the Milky Way halo clusters, the relative 
importance of rotation would increase.

It is interesting to compare the kinematics of the GC systems of NGC~5128
and NGC~4472 (at $\epsilon\sim0.2$ in Figure~\ref{vsig}), 
both of which straddle the $(v_{\phi}/\sigma_{\rm los})^*=0.5$ curve suggestive
of velocity anisotropy. Considering the differences in luminosity 
and environment between these two systems, 
the similarities in their kinematics are quite surprising
($L,V_{\rm N4472}/L,V_{\rm N5128}\sim2$; NGC~4472 resides in the Virgo 
cluster, $\sigma_{\rm Virgo}\sim700$ km s$^{-1}$; 
Binggeli et al.~(1993), NGC~5128 dominates a small group, 
$\sigma_{\rm N5128 group}\sim120$ km s$^{-1}$; van den Bergh (2000)).
Whatever the formation histories of these disparate galaxies
may be, they have clearly achieved quite similar {\it global}
kinematic signatures in their GC systems.

\vbox{
\begin{center}
\leavevmode
\hbox{%
\epsfxsize=8cm
\epsffile{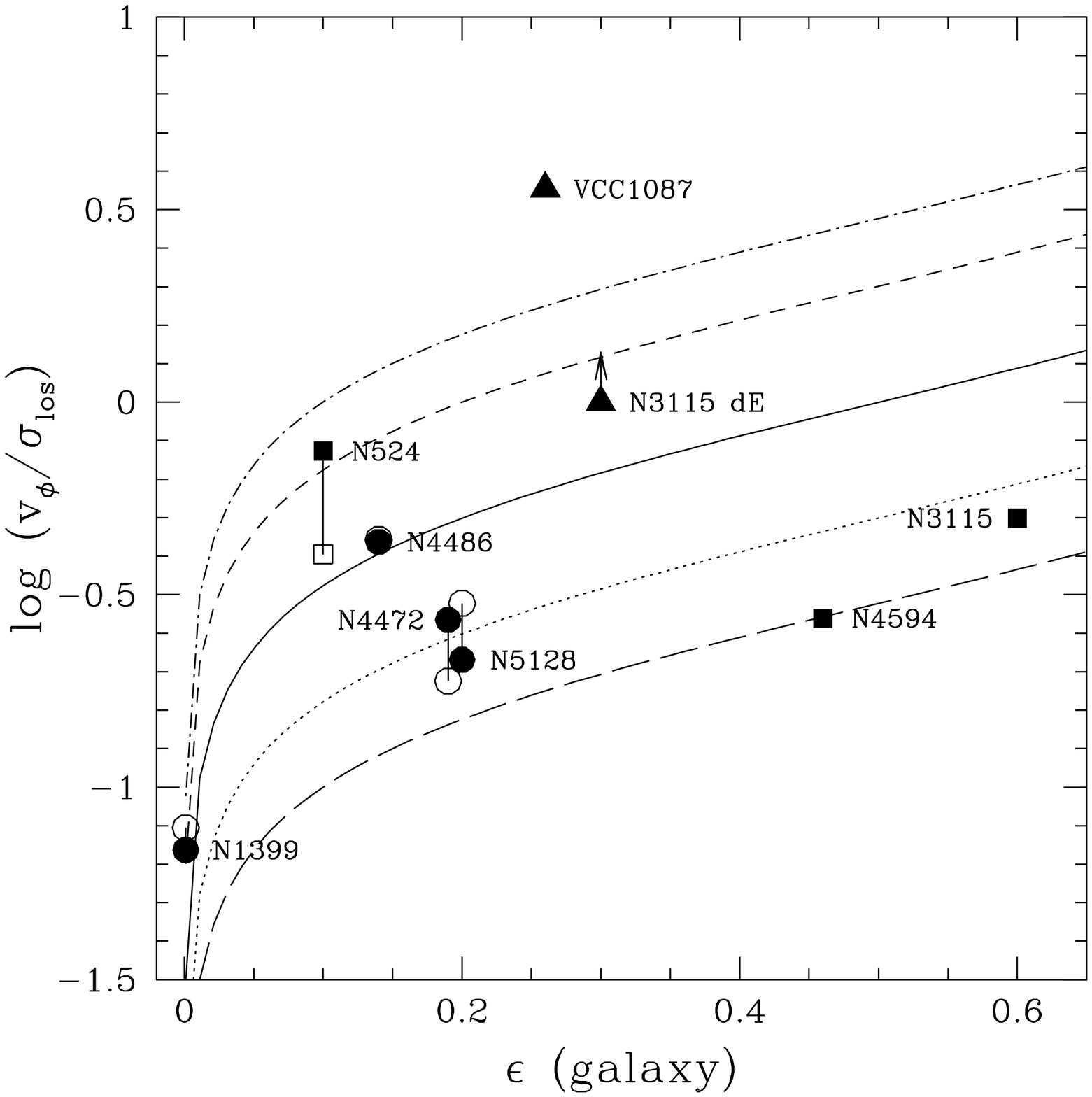}}
\figcaption{\small
The ratio of rotation to velocity dispersion
($v_{\phi}/\sigma_{\rm los}$) of the GC systems
of spheroids versus host galaxy ellipticity.
Connected filled and open circles represent MPC and 
MRC subpopulations, single filled symbols
represent systems in which subpopulations (if any)
have not been separated. 
Circles
indicate elliptical galaxies, squares--S0s, 
triangles--dEs.
The arrow shows NGC~3115 DW1 after
correcting its GC velocity dispersion
for observational errors of 50 km s$^{-1}$.
The solid curve indicates $v_{\phi}/\sigma_{\rm los}$
(Equation~\ref{eq:rot}) for an oblate rotator of ellipticity $\epsilon$
flattened by rotation (i.e., $(v_{\phi}/\sigma_{\rm los})^*=1$).
Curves for $(v_{\phi}/\sigma_{\rm los})^*=$0.3, 0.5, 2 \& 3 
are also shown (long-dashed, dotted, short-dashed and dot-dashed
curves respectively).
\label{vsig}}
\end{center}}

\subsection{Mass-to-Light Ratios}
\label{ML}

GCs represent one of the few probes of the underlying mass distribution 
in gas-poor galaxies beyond the stellar light. 
In fact, they are particularly useful in the case of
dE galaxies which generally have no luminous satellites and 
often have low-surface brightness envelopes which limit the 
utility of studies of their integrated starlight. 

In order to model the radial mass (and mass-to-light ratio, $\Upsilon_V$) profile of VCC~1087, 
we first made the assumption that the VCC~1087 GCs may be regarded 
as an extension of the galaxy starlight. That is to say, we explicitly
assume that the stellar envelope and GCs have the same orbital and spatial distributions.
In so doing, we fit dynamical models to the velocity dispersion
profiles of the galaxy stars (from Geha et al.~2003) as a function of radius
by solving the spherically symmetric Jeans equation 
(see e.g., Geha et al.~2002; Gebhardt \& Fischer 1995; van der Marel 2004):

\begin{equation}
{GM(r) \over r} =  - \sigma_r^2 \left[
{\partial \log \rho \sigma_r^2 \over \partial \log r} + 2 \beta \right] 
\label{Jeans}
\end{equation}

\noindent where $\rho$ is the mass density derived from the galaxy
surface brightness profile and $\sigma_r$ is the radial velocity dispersion.
The $\beta$ parameter describes the velocity 
dispersion anisotropy, which we assume is constant with radius.

\vbox{
\begin{center}
\leavevmode
\hbox{%
\epsfxsize=8cm
\epsffile{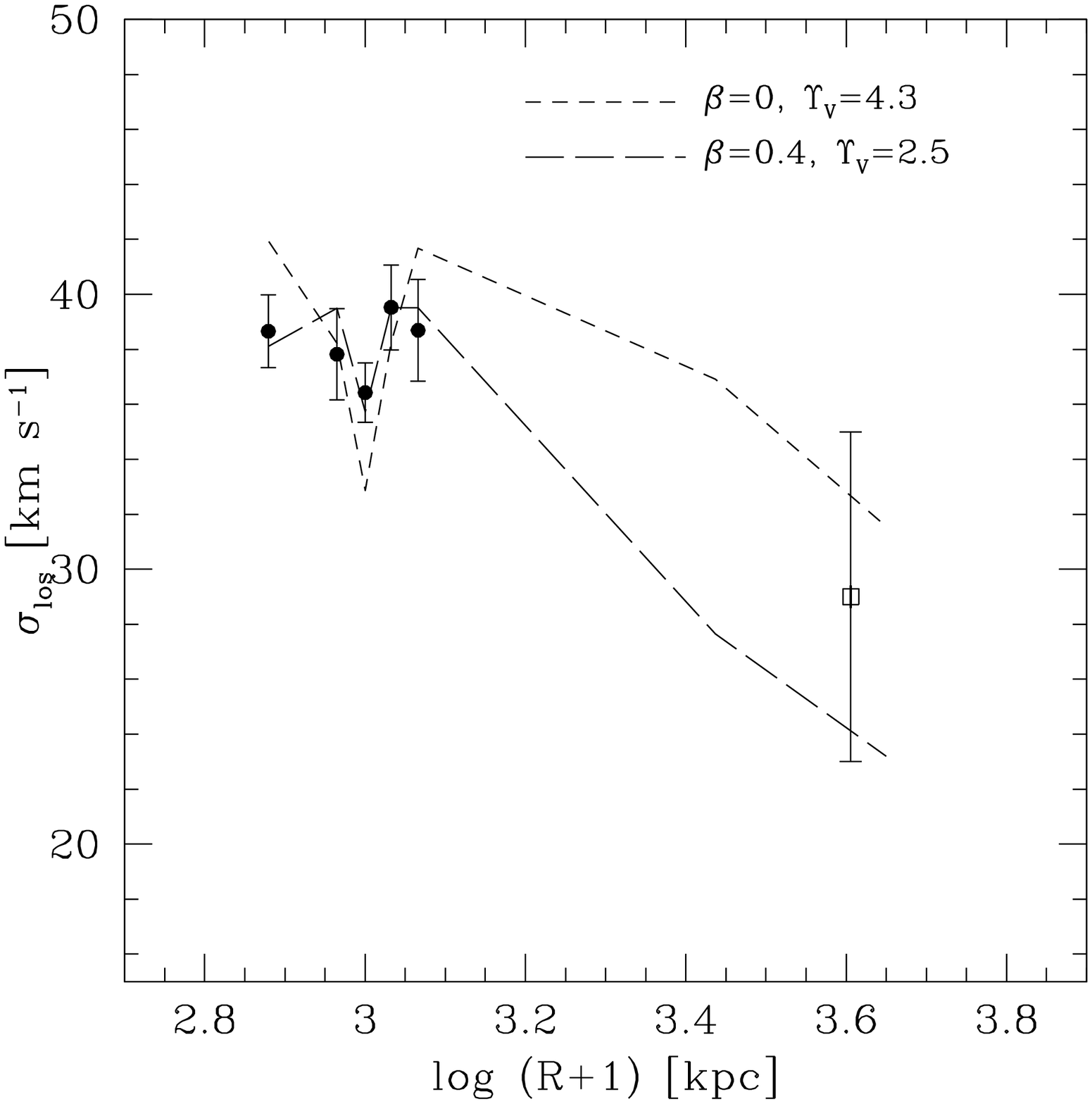}}
\figcaption{\small
Velocity dispersion of the stellar light (solid circles) and 
GCs (open square) of VCC~1087 as a function of galactocentric
radius. Short and long dashed
lines indicate isotropic and anisotropic models of 
constant $\Upsilon_V$
fit to the {\it stellar data only} (see text).
A constant of 10 arcsec has been added to the $x$-axis
for display purposes such that log (R+1 kpc)=3 corresponds to the 
center of VCC~1087.
\label{dispersion}}
\end{center}}

The velocity dispersion of the galaxy stars and GCs 
as a function of galactocentric radius is shown in Figure~\ref{dispersion}.
In the figure, two models of constant $\Upsilon_V$ are shown, 
fit to the stellar data only. One is for an isotropic velocity dispersion ($\beta=0$) and one
is radially anisotropic ($\beta=0.4$). A $\chi^2$ test indicates
that the anisotropic model is slightly preferred over an isotropic
velocity dispersion. In either case, Figure~\ref{dispersion} suggests
$\Upsilon_V\sim3$ as traced by the GCs at a {\it mean} radius of $\sim3$ kpc.

The right-hand side of Equation~\ref{Jeans} only describes the
contribution of ``pressure'' to the total mass of the system.
This is appropriate for the galaxy stars which show no sign
of rotation (Geha et al.~2003), but is incorrect for the GCs
for which rotation appears dynamically significant 
(Section~\ref{Kinematics}). Moreover, with only 12 clusters distributed
over a wide range of radii, the validity of using the Jeans equation 
at all must be questioned. Perhaps more appropriately, we can 
use statistical estimators to estimate the mass traced by the GCs
which will also allow for a check on the above results.

Evans et al.~(2003) introduced the tracer mass estimator (TME), 
which is a generalization of the projected mass estimator
for ``test particles'' (Heisler et al.~1985).
We have used the TME in order to estimate the total gravitating mass 
enclosed within concentric shells out to the projected radius of the outermost GC.
The TME takes the form:

\begin{equation}
M(R<R_{\rm max})=\frac{C}{GN}\sum_i v^2_i R_i w_i
\label{TME}
\end{equation}

\noindent where $N$ is the number of test particles (in this case GCs), 
$v_i$ is the line-of-sight velocity of the $i$th GC and 
$R_i$ is the projected distance of the $i$th GC
from the center of the galaxy. 
Assuming an isothermal-like potential, $C$ is given by (Evans et al.~2003):

\begin{equation}
C=\frac{4\gamma}{\pi}\frac{4-\gamma}{3-\gamma}\frac{1-(r_{\rm in}/r_{\rm out})^{3-\gamma}}
{1-(r_{\rm in}/r_{\rm out})^{4-\gamma}}
\label{TME2}
\end{equation}

\noindent here, $r_{\rm in}$ and $r_{\rm out}$ are the inner and outer radii at which the
population density falls off as $r^{-\gamma}$.
Note that we have included a 
weighting term, $w_i$, in Equation~(\ref{TME}) to prevent undue
influence being given to GCs with large velocity uncertainties, 

\begin{equation}
\sum_i w_i=\sum_i \sigma^{-1}_{v,i}=N
\label{Weights}
\end{equation}

\noindent where $\sigma_{v,i}$ is the velocity uncertainty 
of the $i$th GC.  

We construct three radial bins for which ($r_{\rm in},r_{\rm out}$)
are (0,2), (2,4) \& (4,9) kpc respectively. 
We also set $\gamma=2.1$ for all three bins. The density profile
of the GC appears to flatten in the innermost bin (Figure~\ref{profile}); reducing 
$\gamma$ by 50\% leads to $\sim65\%$ reduction in $\Upsilon_V$.
The ``light'' in $\Upsilon_V$ comes from the $V$-band surface 
brightness profile of Geha et al.~(2003) (Figure~\ref{profile}), 
scaled to integrate to M$_V$=--17.8 (Jerjen et al.~2004; Barazza et al.~2003).

Values for $\Upsilon_V$ estimated using the TME are shown as solid
circles in Figure~\ref{ML}. Using this approach, we find $\Upsilon_V\sim3$ 
which is consistent with that estimated from Equation~\ref{Jeans}.
However, as noted previously, the GC system shows significant rotation.
The curve in Figure~\ref{ML} indicates $\Upsilon_V(r)$, assuming
solid-body rotation in the GC system, with $M(r)=v^2_{\phi}(r) r/G$.
As advocated by Evans et al.~(2003), 
$M_{\rm pressure}$+$M_{\rm rotation}$=$M_{\rm total}$, and our
corresponding values for $\Upsilon_V$ are shown by the open circles
in Figure~\ref{ML}.

The principle uncertainties in the mass estimates stem from the unknown
orbital configuration of the GCs, 
the systemic velocity of the GC system, and the
number of clusters in the sample.
The uncertainty in the distance to VCC~1087 corresponds to errors
in the mass estimates of only $\sim5$\%.
Within these relatively large uncertainties, 
even with the inclusion of $M_{\rm rotation}$, 
$\Upsilon_V$ is flat with radius, and is consistent
with that estimated from the stellar light at small radii.
For a solar metallicity, 5 Gyr old stellar population 
(Section~\ref{Ages}) the Bruzual \& Charlot~(2003) models predict 
$\Upsilon_V$ of 3 (2) for a Salpeter (Chabrier)
initial mass function (IMF). With the assumption of an 
entirely old (12 Gyr), solar metallicity stellar population,
the Bruzual \& Charlot~(2003) predict $\Upsilon_V$=7 (4)
for a Salpeter (Chabrier) IMF. Similar values are found
for the Maraston (1998) stellar population models.
Therefore, $\Upsilon_V$ measured from the GCs out to 
$\sim6.5$ kpc is consistent with that expected for a 
purely baryonic, old stellar population.
Kinematics for a larger number of GCs at large radii
would be extremely valuable for further investigation 
of the mass distribution of VCC~1087.

\vbox{
\begin{center}
\leavevmode
\hbox{%
\epsfxsize=8cm
\epsffile{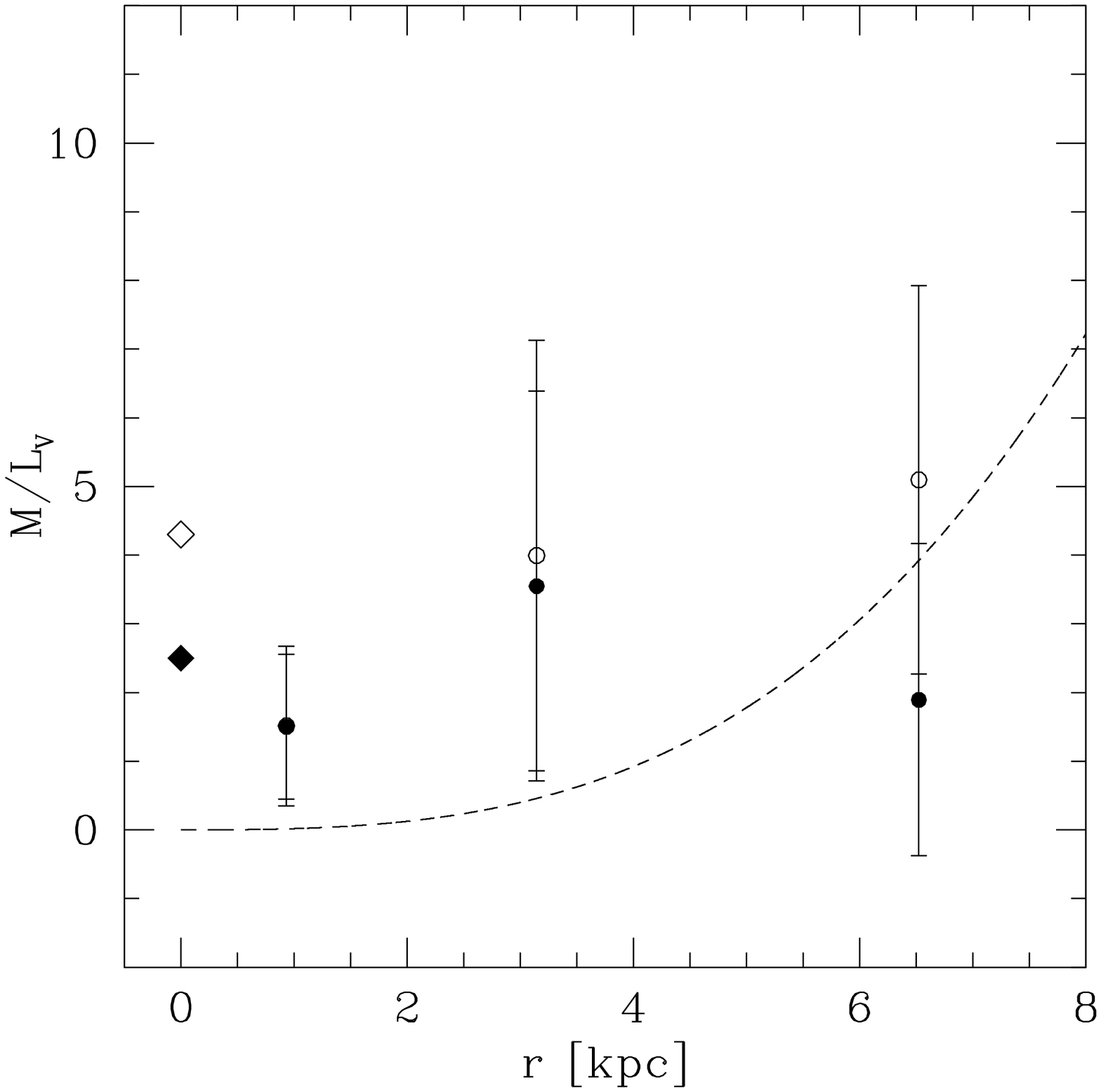}}
\figcaption{\small
Mass-to-light ratio as a function of radius in VCC~1087.
Solid circles are the predictions assuming purely 
pressure support in the tracer mass estimator (Evans et al.~2003), 
open circles include the addition of mass from the rotation
in the GC system (dashed curve). Filled and open diamonds show the 
predictions for the stellar light (Geha et al.~2003) using 
Equation~\ref{Jeans} and assuming an anisotropy parameter ($\beta$)
of 0.4 and 0 respectively (see text).
\label{ML}}
\end{center}}

\subsection{Ages and Abundances}
\label{Ages}

We have investigated the stellar populations of the 
GC system by using Lick indices (Table~\ref{Indices}) in conjunction
with the simple stellar population (SSP) models of 
Thomas, Maraston \& Bender (2003), 
and Thomas, Maraston \& Korn (2004). These models (hereafter
collectively referred to as TMK04), attempt to account for 
non-solar abundance ratios in simple stellar populations by using
synthetic atmosphere models to predict Lick index ``response functions''
for varying abundance ratios. Note that the assumption of an SSP  
probably breaks down for a dE as luminous 
as VCC~1087 (L$_V\sim1.2\times10^9\Lsun$). Therefore, 
the following derived stellar population parameters of VCC~1087 
itself are strictly luminosity-weighted properties only.

In Figure~\ref{HgF}, we show an example index-index 
diagnostic plot of the Lick indices of the GCs compared to the 
TMK04 models. We have chosen the [$\alpha$/Fe]=0.3 
models here (see below). Qualitatively, the GCs generally appear old, and 
have metallicities less than solar.
The galaxy starlight, on the other hand, is more metal-rich, 
and adopting the [$\alpha$/Fe]=0.0 models for this galaxy, 
lies between the 4 and 5 Gyr model lines.
Also shown in Figure~\ref{HgF} are Lick indices 
for 40 Galactic GCs measured from the library of 
Schiavon et al.~(2005). These GCs form a tight sequence, reflecting the high
quality of these data. However, they are clearly offset 
from the models (as are the VCC~1087 data, in the mean), an offset which is significantly larger
than the uncertainties in correcting to the Lick system.
Investigating the origins of this offset (seen in the other
Balmer indices to lesser or greater degrees) is beyond the scope
of this paper, but clearly needs to be resolved 
before confidence can be put into the absolute age-scale of 
SSP models (e.g., Lee \& Worthey 2005).

\vbox{
\begin{center}
\leavevmode
\hbox{%
\epsfxsize=8cm
\epsffile{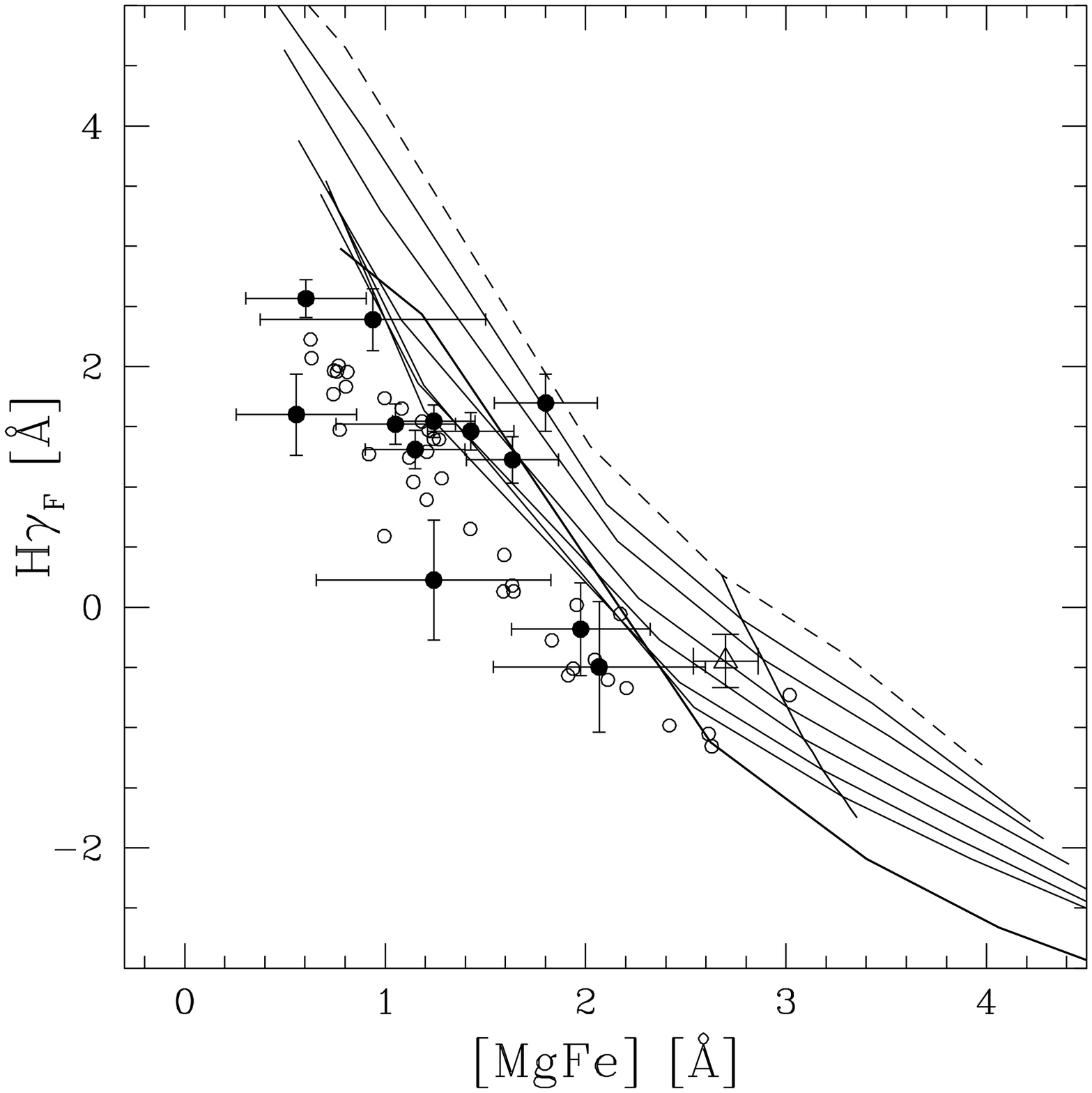}}
\figcaption{\small
The [MgFe]--H$\gamma_{\rm F}$ indices of the VCC~1087 GCs
(solid circles) compared to the stellar population models of TMK04.
Lick indices measured from the Galactic GC sample of 
Schiavon et al.~(2005) are shown as open circles.
The heavy line indicates an age of 15 Gyr, the 
dashed line indicates an age of 3 Gyr. A line
of constant solar metallicity is also shown.
The GCs are appear generally old within the uncertainties, 
VCC~1087 itself (open triangle) appears more metal-rich 
and younger. 
\label{HgF}}
\end{center}}

To place these impressions on a quantitative footing, 
metallicities (here denoted [m/H]), ages and $\alpha$-abundances
(here denoted [E/Fe] -- see Proctor \& Sansom 2002) were derived 
for the GCs and galaxy starlight
from their Lick indices by performing multivariate fits to
the TMK04 models. This technique has been described
extensively in Proctor \& Sansom (2002) and Proctor et al.~(2004).
In general, 10--15 indices were fit simultaneously, and we 
show the resulting best-fit ages and metallicities using this 
procedure in Figure~\ref{MH}.
The same procedure was carried out for the Schiavon et al.~(2005)
data. 
The VCC~1087 GCs exhibit
a range of metallicities ($-1.8\lsim$[m/H]$\lsim-0.8$)
with the three most metal-rich GCs entering into the domain
of the red peak of the Galactic GC system. Their position in
Figure~\ref{MH} is consistent with that shown in Figure~\ref{CMD}, 
i.e., at or near the separation of the two subpopulations.
Note that the Schiavon et al.~library comprises of $\sim25\%$ of the 
known GCs in the Milky Way, whereas our spectroscopic sample
covers the brightest $\sim15\%$ of the VCC~1087 GC system.

Figure~\ref{MH} suggests that the VCC~1087 GCs are all old, 
comparable to the Galactic GCs ages, and within the uncertainties (typically 3 Gyr) 
are coeval. 
By way of contrast, the luminosity-weighted properties
of VCC~1087 itself are quite different. It is metal-rich 
([m/H]$=-0.1$), lying between the metal-rich Galactic
GCs NGC~6528 and NGC~6553 in Figure~\ref{MH}, and intermediate-aged ($\sim4$ Gyr).
This is consistent with the findings of Geha et al.~(2003)
who also used Lick indices and the TMK04 models.

\vbox{
\begin{center}
\leavevmode
\hbox{%
\epsfxsize=8cm
\epsffile{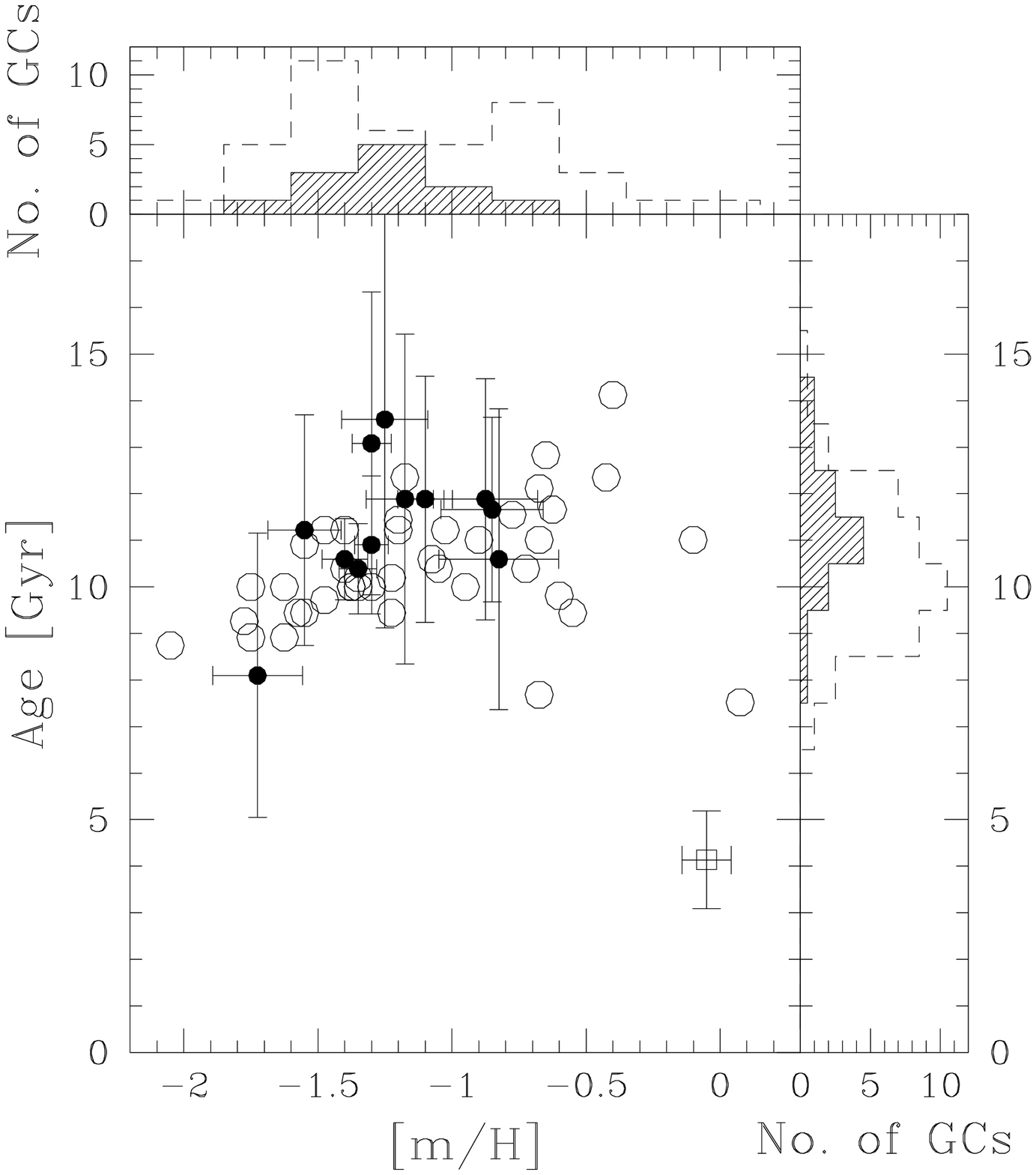}}
\figcaption{\small
Ages and metallicities of the VCC~1087 GCs (solid circles) predicted using the TMK04
stellar population models. 
Also shown is the luminosity-weighted age and
metallicity of the integrated starlight of VCC~1087 (open square), 
and of a sample of 40 Galactic GCs from Schiavon et al.~(2005) (open circles).
The VCC~1087 clusters are old and coeval, with a
range of metallicities.
The shaded and dashed histograms are for
the VCC~1087 and Galactic samples respectively.
\label{MH}}
\end{center}}

Our solutions for the $\alpha$-element abundances
of the VCC~1087 GCs are plotted
in Figure~\ref{EFE}. This is compared to the Galactic GC sample. 
Arguably, the most reliable ``pure'' $\alpha$-capture element
(i.e., that which is least affected by nucleosynthetic evolution
and most easily measured) is Ca (Gratton et al.~2004), and we 
plot the mean [Ca/Fe] ratio from high dispersion analyses in Figure~\ref{EFE}.
For the Galactic GCs, $\langle$[E/Fe]$\rangle\simeq$[Ca/Fe]=0.25, with
dispersions of 0.16 and 0.11 dex respectively.
The VCC~1087 GCs have $\langle$[E/Fe]$\rangle=0.28$ with a dispersion
of 0.16 dex. Among these GCs there appears to be a weak correlation in the sense that 
[E/Fe] decreases with increasing [m/H], but this is does not have statistical significance. 
The position of VCC~1087 in Figure~\ref{EFE}
suggests a solar $\alpha$-abundance, which is consistent
with that found by Geha et al.~(2003).

\vbox{
\begin{center}
\leavevmode
\hbox{%
\epsfxsize=8cm
\epsffile{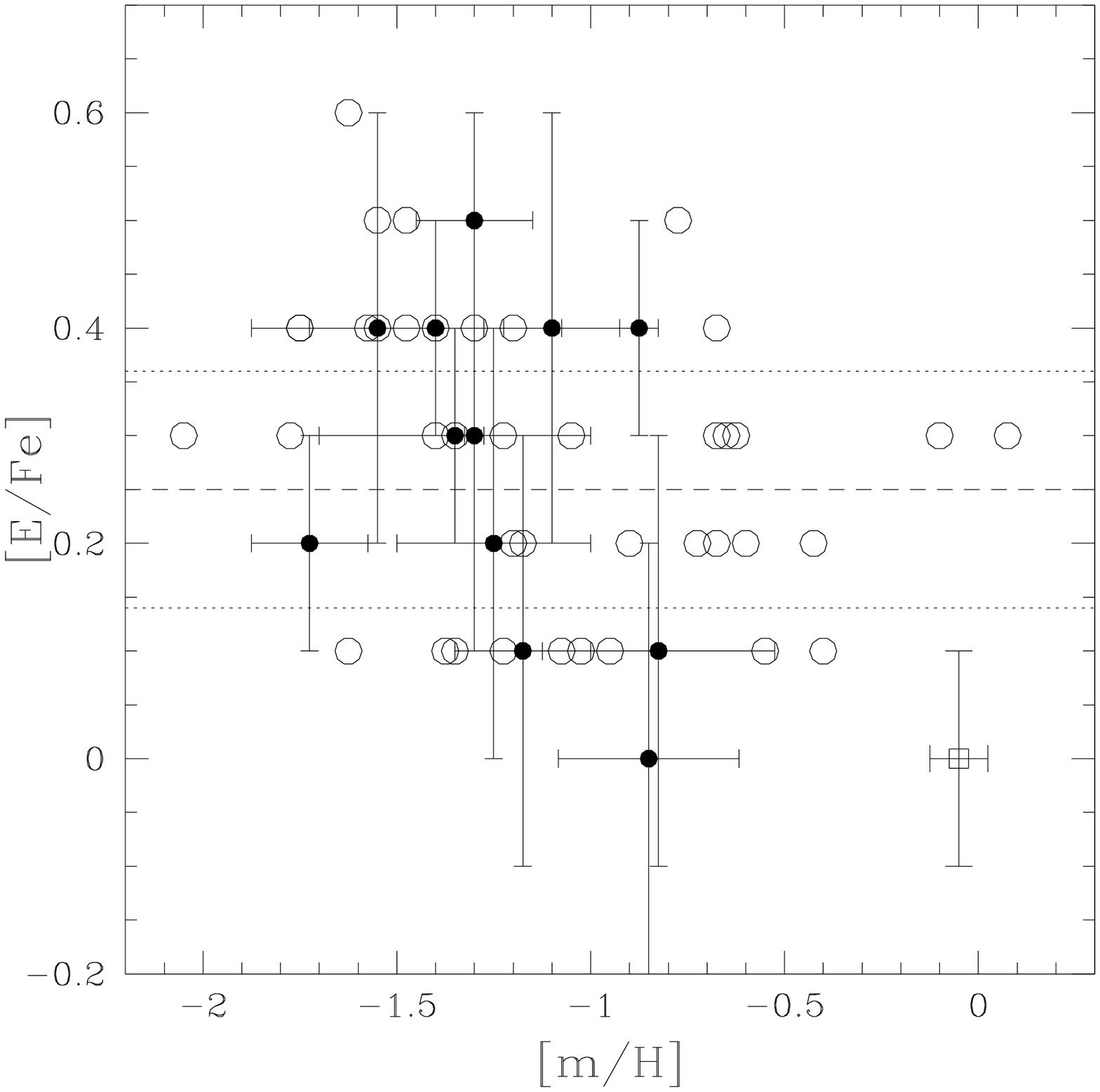}}
\figcaption{\small
Predicted $\alpha$-element abundances of the VCC~1087 GCs
using the TMK04 models (solid circles) and the galaxy starlight (open square)
versus metallicity. 
The Galactic GCs of Schiavon et al.~(2005) are also shown (open circles)
The dashed line (dotted lines are the rms scatter) indicates the mean value of [Ca/Fe] 
for Galactic GCs obtained from high-dispersion analyses (Gratton et al.~2004
and references therein). 
\label{EFE}}
\end{center}}

\section{Discussion}
\label{Discussion}

With M$_V=-17.8$, VCC~1087 lies at the bright
end of the dE luminosity function (e.g., Sandage, Binggeli
\& Tammann 1985).
As such, specific conclusions for this one object are
not necessarily generalizable to dEs as a class.
This caveat notwithstanding, we discuss the 
findings of this study in the context of dE galaxy
formation.

Moore et al.~(1998) and Mastropietro et al.~(2005) 
presented a series of
numerical simulations which demonstrated that 
spiral galaxies can be transformed
into spheroidal galaxies (i.e., dE/dSph types) in the 
cluster environment, ``galaxy harassment''.
The combined effects of the cluster
tidal field and rapid encounters with cluster 
galaxies act to heat infalling progenitor spirals 
into, ultimately, prolate systems.
Could VCC~1087 have been transformed in such a manner, 
and does its GC system offer any clues?

One constraint on this picture is the number of GCs per
unit ($V$-band) luminosity (S$_{\rm N}$).
Miller et al.~(1998) measured S$_{\rm N}$
for 24 dE galaxies in the Virgo and Fornax clusters, and in 
the Leo group, 13 of which are of the dE,N class.
These authors obtained a mean S$_{\rm N}$ of $6.5\pm1.2$, and
found a trend of decreasing S$_{\rm N}$ with increasing M$_V$.	
Miller et al. argued that luminous dE,N galaxies are unlikely to have 
evolved from faded irregular galaxies since
locally irregulars exhibit S$_{\rm N}\sim$0.5--2.
\footnote{Interestingly, Seth et al.~(2004) have
recently found that Virgo and Fornax cluster 
irregulars may have a larger spread in S$_{\rm N}$ than
their field counterparts, with S$_{\rm N}$ ranging from $\sim0$ to 9.
If confirmed, this would allow for the transformation 
of cluster dIrrs to dEs (at least, that is, on the basis of an S$_{\rm N}$ 
argument).}
We estimate a total GC population
of $77\pm19$ GCs in VCC~1087, giving 
S$_{\rm N}=5.8\pm1.4$ (adopting the slightly brighter value of M$_V$=--17.96 
for VCC~1087 (Jerjen et al.~2004), we obtain S$_{\rm N}=5.1\pm1.2$).
These values are significantly higher than the typical
S$_{\rm N}\sim1$ found for late-type
spirals (e.g., Olsen et al.~2004; 
Chandar et al.~2004).  In order to harbor $\sim80$ GCs, a progenitor S$_{\rm N}=1$ spiral
galaxy would need to have had M$_V\sim-19.8$.
Any progenitor spiral must also have been essentially
``pure disk'', since the densities of spiral
bulges are too high to be strongly effected by 
harassment (Moore et al.~1998).

The present-day Galaxy luminosity function 
in the Virgo cluster shows that, for 
M$_V\sim-20$, Irr galaxies are not represented, and Sd galaxies
barely so (Binggeli et al. 1988). However, the peak of a Gaussian luminosity
function for Sc galaxies does occur near this 
magnitude.
An Sc galaxy which fell into the Virgo cluster $\sim5$ Gyr ago
(e.g., Conselice et al.~2001), and 
consequently lost its gas (via ram-pressure stripping; 
Gunn \& Gott 1972), 
could plausibly fade by 1--2 mag in the 
$V$-band. Subsequent mass loss through
tidal encounters could fade the galaxy further, although
this would presumably result in GCs being stripped
also, lowering the S$_{\rm N}$ of the galaxy.
Nevertheless, in terms of GC numbers, a faded
Sc spiral could approach the 
S$_{\rm N}$ of VCC~1087. 

By the same
token, the high S$_{\rm N}$ of VCC~1087
is inconsistent with the characteristic
S$_{\rm N}\sim2$ observed for low-luminosity
ellipticals (e.g., Ashman \& Zepf 1998).
Admittedly, data is currently scarce 
on the S$_{\rm N}$ of such galaxies, 
particularly regarding the question of 
the variation of S$_{\rm N}$ with environment. However, the 
present observations sit uncomfortably with the notion 
that VCC~1087 may be simply at the faint end
of low-luminosity elliptical galaxies (c.f., Miller et al.~1998).

In principle, stellar kinematics provide constraints
on dE formation models. Simulations suggest
that galaxy harassment is more efficient at
increasing velocity dispersions (through
bar formation and the transformation of
circular to radial orbits) than disrupting 
rotation (Moore et al.~1998).
Consequently, $(v_{\rm rot}/\sigma_{\rm los})^*$
should decrease due to the harassment 
process, but will remain non-zero except in cases
where the progenitor spirals are strongly disrupted
near to the cluster center
(Mastropietro et al.~2005). Such severe disruption 
which may lead to non-rotating dEs is expected to be reflected
in their observed morphologies and orbital characteristics.
Observationally, Geha et al.~(2003) found no obvious 
differences between the stellar populations of 
a sample 13 non-rotating and 4 rotating dEs.
Moreover, in the specific case of VCC~1087, 
Geha et al.~(2003) found no evidence for rotation.

This seems to present something of a challenge
to the theoretical models.
However, as discussed by Mastropietro et al.~(2005), 
kinematical observations of dE galaxies are fairly
sparse, and not without controversy.
There are five galaxies in common
between Geha et al.~(2003) and the kinematical study 
of van Zee et al.~(2004b). The latter study generally 
found higher maximum rotation velocities. Mastropietro et al.
argue that this difference has arisen due to the
fact that the van Zee et al.~(2004b) observations covered a greater 
radial extent than those of Geha et al.~(2003). Clearly
further studies are required. 

In terms of GC kinematics the predictions are less
clear since, to date, simulations have not explicitly taken
them into account. Presumably, they will behave similarly
to the dissipationless stellar component of the 
harassed spirals if they have the same density
profile as the galaxy.
Under the assumption that the azimuthal spatial distributions of GCs
follows that of the galaxy stars, the GC systems
of spheroids generally possess a significant degree
of pressure support (Figure~\ref{vsig}).
However, the GC system of VCC~1087 appears to be 
rotationally flattened, which is possibly also the 
case for the only other GC system of a luminous dE
studied to date, NGC~3115 DW1 (Puzia et al.~2000).
In contrast, the old halo Milky Way GCs, of which the MPCs
in VCC~1087 are presumably analogs, show no 
significant rotation. We note
that if in fact the VCC~1087 GCs represent the vestiges of a disk
system, their rotation (i.e., an `H\small{I} line-width'
of $\sim210$ km s$^{-1}$) would place such a disk 
on the $V$-band Tully-Fisher relation at $M_V\sim$--18.7
(Sakai et al.~2000). If it is subsequently shown that the rotation 
in the GCs persists out to larger radii, this would 
correspondingly increase this magnitude estimate.

The nucleus of VCC~1087 has a half-light radius ($\sim3$~pc) 
similar to the GCs, but appears $\sim$ 0.1 mag redder.
Tremaine, Ostriker \& Spitzer (1975) first suggested
that galaxy nuclei (in their case, that of M31) may be
built from the merged remnants of massive GCs whose orbits decayed
by dynamical friction. This idea becomes particularly
appealing considering the evidence for a deficit of 
GCs, or a ``core'', in the GC 
spatial density profiles in the inner regions of dE
galaxies (Durrell et al.~1996a; Lotz et al.~2001; 
Section~\ref{Photometry}).
Using Monte Carlo simulations, 
Lotz et al.~(2001) found that the brightest nuclei
of dE galaxies (M$_V\sim-12$) may well have been
formed in such a manner, but the nuclei 
of fainter dEs were several magnitudes fainter than predicted
(see also Miller et al.~1998). For VCC~1087, we obtain M$_V\sim-11.4$, 
which is a borderline case in the Lotz et al.~(2001)
simulations (see their figure 8).

However, the color of the nucleus presents something of a problem
for this scenario. As mentioned above, 
the nucleus has a color consistent with, or slightly redder
than the putative MRCs. This would suggest that 
the nucleus is {\it solely} constructed
from decayed MRCs. This seems somewhat unlikely, 
considering that the VCC~1087 GC system consists
predominantly of MPCs. It is possible that whatever 
lead to the formation of the MRCs formed the majority
with a very high central concentration, leading to a 
shorter timescale for dynamical friction.
Numerical simulations would be useful
for investigating this issue.
Interestingly, the Moore et al.~(1998) 
simulations with gas also predict the existence of a possible nucleus.
Gas sinks to the center of the galaxy in their simulations due to the 
combination of tidal torques and cooling, 
leading to a central density excess in their model galaxies.

From the previous considerations, a plausible evolutionary
scenario for VCC~1087 may have proceeded as follows:
an Sc spiral with M$_V\sim-20$ and $\sim$100 GCs
fell into the Virgo cluster approximately 5 Gyr ago
(the luminosity-weighted age of VCC~1087).
Gas removal through ram-pressure truncated further
star formation, and the galaxy stellar populations
passively evolved and faded by $\sim$1.5 mag.
Subsequent tidal interactions
with cluster members and the cluster tidal field perturbed
and heated the galaxy, disrupting the disk and forming
a low surface brightness spheroid. Possibly, some field
stars and GCs where also stripped during this process.
Any remaining gas in the system lost angular momentum
from torques and cooled, leading to star formation
in the central regions creating a solar metallicity
nucleus, embedded in a diffuse envelope with a retinue
of surviving GCs.

\section{Summary}
\label{Summary}

We have performed a photometric and 
spectroscopic analysis of the GC system
of the Virgo dE galaxy VCC~1087. This is the first 
such study of a dE galaxy in a cluster environment.
A total of 68 GCs are detected in {\it HST}/ACS imaging
to $\g\sim26.5$. Correcting for areal incompleteness and for the faint
end of the luminosity function this gives a total
GC population of $77\pm19$ GCs, corresponding to S$_{\rm N}=5.8\pm1.4$
in the $V$-band.
The color-magnitude diagram of the GCs reveals
a pronounced blue peak ($\gz\sim0.90)$ and 
a tail of red GCs ($\gz\sim1.17$). Thus, VCC~1087
may have a metal-rich population of GCs similar
to those seen in luminous ellipticals.
Although poorly constrained due to small numbers, the 
spatial density profile of the GCs is surprisingly
shallow, with a power-law exponent of $\alpha=-1.1\pm0.4$.
This is similar to the spatial distribution of GCs seen in cD galaxies, 
rather than those in low-luminosity ellipticals.
The GC luminosity function (GCLF) appears log-normal, 
and has a turn-over magnitude very similar to those
in luminous Virgo ellipticals (Strader et al.~2005).

We investigated the kinematics and stellar populations
of a subset of 12 GCs using Keck/LRIS spectroscopy.
The GCs show significant rotation (V$_{\phi}=104\pm35$ km s$^{-1}$).
With a spatially-averaged velocity dispersion of 
$\sigma_{\rm los}=29\pm10$ km s$^{-1}$, this suggests
that this rotation is dynamically significant 
with $(v_{\rm rot}/\sigma_{\rm los})^*>1$.
Comparing existing kinematical data 
for GC systems in early type galaxies 
suggests that the VCC~1087 GCs
are among the most rotationally dominated.
Considering both the Jeans equation and the tracer mass estimator (Evans et al.~2003)
we derive $V$--band mass-to-light ratios ($\Upsilon_V$)
for the galaxy in the range $1 \leq \Upsilon_V \leq 8$ within 
$\sim6.5$ kpc of the galaxy center.
These values are consistent with $\Upsilon_V$
of the galaxy stars, and can be understood purely in terms
of a baryonic old, metal-rich stellar population. 
Stellar population analysis using Lick
indices shows that the GCs are all old ($\gsim10$ Gyr), 
with a range of metallicities.
We also estimate [$\alpha$/Fe]$\sim$0.3, 
similar to the Galactic GC system.
In contrast, the luminosity-weighted galaxy starlight
has approximately solar metallicity, solar
[$\alpha$/Fe], and is intermediate-aged ($\sim4$ Gyr).

We discuss the characteristics of VCC~1087 and its GC 
system in the context of the galaxy ``harassment'' scenario
(e.g., Moore et al. 1998). We conclude that
the observations are consistent with VCC~1087 being the 
remains of a faded, tidally perturbed Sc spiral, although this 
is probably not a unique interpretation of these data.

\section{Acknowledgments}

Funding support comes from NSF grant AST 02-06139.
M.~G.~is supported by NASA through Hubble Fellowship grant
HF-01159.01-A awarded by the Space Telescope Science Institute, which
is operated by the Association of Universities for Research in
Astronomy, Inc., under NASA contract NAS 5-26555.
We thank Lee Spitler for assistance in the reduction of the {\it HST}/ACS data, and the
anonymous referee who provided an extremely constructive critique of the manuscript. 
The authors wish to recognize and acknowledge the very significant cultural role and 
reverence that the summit of Mauna Kea has always had within the indigenous Hawaiian 
community. We are most fortunate to have the opportunity to conduct observations from this mountain.

\clearpage

\begin{deluxetable}{lcclllc}
\renewcommand{\arraystretch}{.6} 
\footnotesize
\tablecaption{Basic Data for VCC~1087 Globular Clusters \label{tab1}}
\tablewidth{0pt}
\tablehead{
\colhead{ID} & \colhead{$\alpha$(J2000)}   & \colhead{$\delta$(J2000)}   &
\colhead{$\g$} & \colhead{$\z$}  & \colhead{$\gz$} & \colhead{Velocity} \\
\colhead{} & \colhead{(degrees)}  & \colhead{(degrees)} & \colhead{(AB mag)} & 
\colhead{(AB mag)} & \colhead{(AB mag)} & \colhead{(km s$^{-1}$)} \\ 
}
\startdata
GC1 & 187.05648 & 11.78853 & 21.885 $\pm$ 0.007& 20.966 $\pm$ 0.007& 0.919 $\pm$ 0.001 & 608 $\pm$ 12 \\
GC2 & 187.07696 & 11.77774 & 21.950 $\pm$ 0.008& 21.141 $\pm$ 0.008& 0.809 $\pm$ 0.011 & 783 $\pm$ 9 \\
GC3 & 187.06023 & 11.79187 & 22.176 $\pm$ 0.011& 21.305 $\pm$ 0.014& 0.871 $\pm$ 0.018 & 701 $\pm$ 22 \\
GC4 & 187.05713 & 11.77568 & 22.209 $\pm$ 0.008& 21.408 $\pm$ 0.008& 0.801 $\pm$ 0.011 & 695 $\pm$ 14 \\
GC5 & 187.05296 & 11.78424 & 22.342 $\pm$ 0.009& 21.485 $\pm$ 0.009& 0.857 $\pm$ 0.012 & 665 $\pm$ 10 \\
GC6 & 187.06281 & 11.79148 & 22.397 $\pm$ 0.017& 21.550 $\pm$ 0.025& 0.847 $\pm$ 0.030 & 647 $\pm$ 10 \\
GC9 & 187.06194 & 11.78850 & 22.955 $\pm$ 0.035& 21.892 $\pm$ 0.044& 1.063 $\pm$ 0.056 & 665 $\pm$ 35 \\
GC14 & 187.08450 & 11.80643 & 22.892 $\pm$ 0.014& 22.008 $\pm$ 0.032& 0.884 $\pm$ 0.035 & 745 $\pm$ 13 \\
GC16 & 187.06472 & 11.79208 & 22.990 $\pm$ 0.016& 22.123 $\pm$ 0.017& 0.867 $\pm$ 0.023 & 701 $\pm$ 34 \\
GC22 & 187.06795 & 11.78524 & 23.440 $\pm$ 0.017& 22.359 $\pm$ 0.015& 1.081 $\pm$ 0.022 & 631 $\pm$ 49 \\
GC25 & 187.07155 & 11.78644 & 22.299 $\pm$ 0.009& 21.378 $\pm$ 0.008& 0.921 $\pm$ 0.012 & 717 $\pm$ 31 \\
GC26 & 187.07288 & 11.78654 & 23.713 $\pm$ 0.021& 22.584 $\pm$ 0.018& 1.129 $\pm$ 0.027 & 698 $\pm$ 17 \\
Nucleus & 187.06185 & 11.78996 & 20.216 $\pm$ 0.003 & 18.928 $\pm$ 0.003 & 1.295 $\pm$ 0.004 & 646 $\pm$ 30$^a$ \\   
\enddata
\tablenotetext{a}{Velocity from integrated light of nucleus+stellar envelope.}
\end{deluxetable}

\clearpage

\begin{deluxetable}{lcccl}
\renewcommand{\arraystretch}{.6} 
\footnotesize
\tablecaption{Kinematics of GC Systems\label{kin}}
\tablewidth{0pt}
\tablehead{
\colhead{ID} & \colhead{$(v_{\phi}/\sigma_{\rm los})_{\rm MPCs}$}   & \colhead{$(v_{\phi}/\sigma_{\rm los})_{\rm MRCs}$}   
& \colhead{Ellipticity} & \colhead{Source}\\
}
\startdata
NGC~524  &     0.75      & 0.40        &0.10 & Beasley et al.~2004\\
VCC~1087 &     3.60      &...	    &0.26 & this work.\\
NGC~1399 &     0.07      & 0.07        &0.00 & Richtler et al.~2004\\
NGC~3115 &     0.50      &...	    &0.60 & Kuntschner et al.~2002\\
NGC~3115 DW1&  1.00~(1.36)$^a$ &...	    &0.30 & Puzia et al.~2000\\
NGC~4472 &     0.27      & 0.19        &0.19 & C{\^ o}t{\' e} et al.~2003; Zepf et al.~2000\\
NGC~4486 &     0.43      & 0.44        &0.14 & C{\^ o}t{\' e} et al.~2001\\
NGC~4594 &     0.27      &...	    &0.46 & Bridges et al.~1997\\
NGC~5128 &     0.21      & 0.30        &0.20 & Peng et al.~2004\\
\enddata
\tablenotetext{a}{Corrected for velocity uncertainties of 50 km s$^{-1}$ and
using the five inner cluster velocities given in Puzia et al.~(2000).}\end{deluxetable}

\begin{deluxetable}{lrrrrrrrrrrr}
\renewcommand{\arraystretch}{.6} 
\footnotesize
\tablecaption{Lick/IDS indices and uncertainties for VCC~1087 globular clusters.
\label{Indices}}
\tablewidth{0pt}
\tablehead{
\colhead{ID} & \colhead{H$\delta_{\rm A}$}   & \colhead{H$\delta_{\rm F}$}   &
\colhead{CN$_1$} & \colhead{CN$_2$}  & \colhead{Ca4227} &
\colhead{G4300} & \colhead{H$\gamma_{\rm A}$}  & \colhead{H$\gamma_{\rm F}$} &
\colhead{Fe4384} & \colhead{Ca4455}  & \colhead{Fe4531}\\
\colhead{} & \colhead{[\AA]} &  \colhead{[\AA]} & \colhead{[mag]} &  \colhead{[mag]} &
\colhead{[\AA]} & \colhead{[\AA]} &  \colhead{[\AA]} & \colhead{[\AA]} &  \colhead{[\AA]} &
\colhead{[\AA]} & \colhead{[\AA]} \\
}
\startdata
GC1	& 4.07 & 2.50 &--0.102 &--0.072 & 0.28 & 2.94 & 0.05 & 1.31 & 0.32 & 0.32 & 1.96\\
  $\pm$ & 0.24 & 0.17 & 0.007 & 0.009 & 0.14 & 0.24 & 0.25 & 0.16 & 0.37 & 0.19 & 0.29\\
GC2	& 2.94 & 2.62 &--0.039 &--0.026 & 0.46 & 1.59 &--0.23 & 1.55 & 3.79 & 0.14 & 0.28\\
    ... & 0.21 & 0.14 & 0.006 & 0.008 & 0.11 & 0.20 & 0.22 & 0.14 & 0.31 & 0.17 & 0.25\\
GC3	& 1.61 & 1.79 &--0.021 & 0.005 & 0.51 & 3.49 &--0.18 & 1.46 & 1.86 & 0.41 & 1.25\\
    ... & 0.24 & 0.16 & 0.007 & 0.009 & 0.13 & 0.24 & 0.25 & 0.16 & 0.36 & 0.19 & 0.30\\
GC4	& 3.18 & 2.63 &--0.089 &--0.063 & 0.10 & 0.85 & 2.31 & 2.57 & 0.09 & 0.25 & 0.72\\
    ... & 0.25 & 0.17 & 0.007 & 0.009 & 0.13 & 0.26 & 0.25 & 0.16 & 0.40 & 0.20 & 0.29\\
GC5	& 2.82 & 2.85 &--0.081 &--0.037 &--0.27 & 2.14 & 0.80 & 1.52 & 0.03 & 0.92 & 2.25\\
    ... & 0.28 & 0.19 & 0.009 & 0.010 & 0.16 & 0.29 & 0.28 & 0.17 & 0.44 & 0.22 & 0.32\\
GC6	& 2.25 & 2.01 &--0.059 &--0.026 & 0.38 & 2.67 &--0.49 & 1.23 & 1.96 & 0.59 & 1.57\\
    ... & 0.32 & 0.22 & 0.009 & 0.011 & 0.17 & 0.31 & 0.32 & 0.19 & 0.45 & 0.24 & 0.38\\
GC9	&--0.74 & 0.87 & 0.037 & 0.027 &--0.58 & 3.49 &--1.07 & 0.23 &--0.76 & 0.82 & 0.87\\
    ... & 0.56 & 0.31 & 0.020 & 0.023 & 0.37 & 0.67 & 0.76 & 0.50 & 1.23 & 0.66 & 0.96\\
GC14	& 4.75 & 2.60 &--0.103 &--0.066 & 0.45 & 2.89 & 0.70 & 1.70 & 0.94 & 0.71 & 1.85\\
    ... & 0.39 & 0.29 & 0.012 & 0.015 & 0.23 & 0.38 & 0.40 & 0.24 & 0.59 & 0.29 & 0.45\\
GC16	& 2.92 & 2.13 &--0.089 &--0.049 & 0.03 & 4.27 &--0.35 & 1.60 & 0.54 &--0.84 & 0.65\\
    ... & 0.53 & 0.38 & 0.015 & 0.019 & 0.30 & 0.52 & 0.56 & 0.34 & 0.69 & 0.43 & 0.67\\
GC22	& 0.69 & 0.86 &--0.105 &--0.116 & 0.92 &--0.10 & 0.08 &--0.18 & 1.80 & 0.66 & 0.78\\
    ... & 0.66 & 0.45 & 0.019 & 0.023 & 0.29 & 0.71 & 0.63 & 0.38 & 0.92 & 0.47 & 0.69\\
GC25	&--1.56 & 1.57 &--0.024 &--0.011 & 0.04 & 7.06 &--6.64 &--0.50 & 5.62 & 1.61 & 0.99\\
    ... & 0.92 & 0.56 & 0.027 & 0.033 & 0.49 & 0.67 & 0.93 & 0.54 & 1.13 & 0.64 & 1.01\\
GC26	& 0.53 & 0.79 &--0.066 &--0.038 &--0.10 & 3.89 & 1.11 & 2.39 & 0.90 & 0.40 &--0.51\\
    ... & 0.42 & 0.29 & 0.012 & 0.015 & 0.23 & 0.41 & 0.43 & 0.26 & 0.62 & 0.33 & 0.54\\
Galaxy	&--1.28 & 0.85 &--0.046 &--0.023 & 1.69 & 5.03 &--3.65 &--0.45 & 5.24 & 2.05 & 3.25\\
    ... & 0.36 & 0.24 & 0.010 & 0.011 & 0.17 & 0.31 & 0.36 & 0.22 & 0.45 & 0.23 & 0.36\\
\enddata

\end{deluxetable}

\clearpage

\begin{deluxetable}{lrrrrrrrrr} 
\renewcommand{\arraystretch}{.6} 
\footnotesize
\tablecaption{Lick/IDS indices and uncertainties for VCC~1087 globular clusters} 
\tablewidth{0pt}
\tablehead{
\colhead{ID} & \colhead{C$_2$4668} & \colhead{H$\beta$}  & \colhead{Fe5015} & \colhead{Mg$_1$} & \colhead{Mg$_2$}  & 
\colhead{Mg $b$}  & \colhead{Fe5270} & \colhead{Fe5335} & \colhead{Fe5406} \\
\colhead{} & \colhead{[\AA]} & \colhead{[\AA]} & \colhead{[\AA]} & \colhead{[mag]} &  \colhead{[mag]} & 
\colhead{[\AA]} & \colhead{[\AA]} &  \colhead{[\AA]} & \colhead{[\AA]}\\
}
\startdata
GC1	& 0.80 & 2.07 & 2.59 & 0.049 & 0.083 & 0.83 & 1.18 & 0.97 & 0.45\\
  $\pm$ & 0.44 & 0.16 & 0.37 & 0.004 & 0.004 & 0.18 & 0.20 & 0.23 & 0.16\\
GC2	&--2.81 & 2.50 & 1.92 & 0.019 & 0.057 & 0.88 & 1.23 & 1.18 & 0.24\\
    ... & 0.39 & 0.14 & 0.33 & 0.003 & 0.004 & 0.17 & 0.18 & 0.20 & 0.16\\
GC3	& 1.27 & 2.11 & 2.16 & 0.037 & 0.112 & 1.37 & 1.20 & 1.10 & 0.57\\
    ... & 0.41 & 0.17 & 0.35 & 0.004 & 0.005 & 0.19 & 0.19 & 0.24 & 0.16\\
GC4	&--0.23 & 2.79 & 1.79 & 0.011 & 0.025 & 0.07 & 0.76 & 0.79 & 0.49\\
    ... & 0.48 & 0.18 & 0.36 & 0.004 & 0.005 & 0.20 & 0.23 & 0.25 & 0.19\\
GC5	&--0.04 & 2.49 & 1.25 & 0.030 & 0.085 & 0.59 & 1.37 & 0.88 & 0.45\\
    ... & 0.48 & 0.18 & 0.40 & 0.004 & 0.005 & 0.20 & 0.22 & 0.26 & 0.20\\
GC6	& 0.79 & 2.28 & 2.42 & 0.030 & 0.096 & 1.29 & 1.79 & 1.37 & 1.15\\
    ... & 0.55 & 0.21 & 0.47 & 0.005 & 0.006 & 0.23 & 0.26 & 0.32 & 0.22\\
GC9	&--0.86 & 2.44 & 0.57 & 0.031 & 0.134 & 1.12 & 1.43 & 0.61 &--0.19\\
    ... & 1.33 & 0.50 & 1.05 & 0.010 & 0.011 & 0.42 & 0.49 & 0.56 & 0.43\\
GC14	&--0.56 & 2.86 & 1.06 & 0.024 & 0.093 & 1.62 & 1.32 & 1.89 & 0.48\\
    ... & 0.70 & 0.26 & 0.58 & 0.006 & 0.007 & 0.29 & 0.31 & 0.36 & 0.27\\
GC16	& 1.03 & 1.84 & 2.04 & 0.031 & 0.048 & 0.16 &--0.21 & 1.32 &--0.18\\
    ... & 1.01 & 0.39 & 0.86 & 0.009 & 0.010 & 0.42 & 0.47 & 0.53 & 0.41\\
GC22	&--0.14 & 1.47 & 1.70 & 0.027 & 0.089 & 1.44 & 1.39 & 2.86 & 1.16\\
    ... & 1.08 & 0.40 & 0.85 & 0.009 & 0.011 & 0.46 & 0.50 & 0.52 & 0.39\\
GC25	& 0.49 & 1.76 & 3.61 & 0.017 & 0.099 & 2.54 & 1.41 & 1.49 & 1.27\\
    ... & 1.53 & 0.58 & 1.17 & 0.014 & 0.016 & 0.63 & 0.71 & 0.80 & 0.57\\
GC26	& 1.48 & 2.11 & 1.02 & 0.020 & 0.069 & 0.41 & 1.13 & 1.05 &--0.04\\
    ... & 0.71 & 0.32 & 0.65 & 0.007 & 0.008 & 0.34 & 0.38 & 0.42 & 0.30\\
Galaxy	& 2.70 & 2.07 & 5.31 & 0.069 & 0.178 & 2.82 & 3.06 & 2.10 & 1.59\\
    ... & 0.55 & 0.21 & 0.45 & 0.005 & 0.006 & 0.22 & 0.24 & 0.27 & 0.20\\
\enddata
\end{deluxetable}


\begin{references}
\reference{} Ashman, K.~M., \& Zepf, S.~E.\ 1998, Globular cluster systems, Cambridge University Press
\reference{} Ashman, K.~M., Bird, C.M., \& Zepf, S.E., 1994, \aj, 108, 2348
\reference{} Barazza, F.~D., Binggeli, B., \& Jerjen, H.\ 2003, \aap, 407, 121 
\reference{} Barazza, F.~D., Binggeli, B., \& Jerjen, H.\ 2002, \aap, 391, 823 
\reference{} Beasley, M.~A., Forbes, D.~A., Brodie, J.~P., \& Kissler-Patig, M.\ 2004, \mnras, 347, 1150
\reference{} Beers, T.~C., Flynn, K., \& Gebhardt, K.\ 1990, \aj, 100, 32 
\reference{} Borkova, T.~V., \& Marsakov, V.~A.\ 2000, Astronomy Reports, 44, 665 
\reference{} Binggeli, B., Sandage, A., \& Tammann, G.~A.\ 1988, ARA\&A, 26, 509
\reference{} Binney, J.\ 1978, \mnras, 183, 501
\reference{} Bridges, T.~J., Ashman, K.~M., Zepf, S.~E., Carter, D., Hanes, D.~A., Sharples, R.~M., \& Kavelaars, J.~J.\ 1997, \mnras, 284, 376 
\reference{} Caldwell, N., \& Bothun, G.~D.\ 1987, \aj, 94, 1126 
\reference{} Cardelli, J.~A., Clayton, G.~C., \& Mathis, J.~S.\ 1989, \apj, 345, 245 
\reference{} Chandar, R., Whitmore, B., \& Lee, M.~G.\ 2004, \apj, 611, 220 
\reference{} Chandrasekhar, S.\ 1943, \apj, 97, 255
\reference{} Conselice, C.~J.\ 2004, IAU Symposium, 217, 556 
\reference{} Conselice, C.~J., Gallagher, J.~S., \& Wyse, R.~F.~G.\ 2001, \apj, 559, 791
\reference{} C{\^ o}t{\' e}, P., et al.\ 2004, \apjs, 153, 223
\reference{} C{\^ o}t{\' e}, P., McLaughlin, D.~E., Cohen, J.~G., \& Blakeslee, J.~P.\ 2003, \apj, 591, 850
\reference{} C{\^ o}t{\' e}, P., et al.\ 2001, \apj, 559, 828
\reference{} De Propris, R., Phillipps, S., Drinkwater, M.~J., Gregg, M.~D., Jones, J.~B., Evstigneeva, E., \& Bekki, K.\ 2005, \apjl, 623, L105 
\reference{} De Rijcke, S., Dejonghe, H., Zeilinger, W.~W., \& Hau, G.~K.~T.\ 2003, \aap, 400, 119
\reference{} De Rijcke, S., Dejonghe, H., Zeilinger, W.~W., \& Hau, G.~K.~T.\ 2001, \apjl, 559, L21  
\reference{} Dirsch, B., Schuberth, Y., \& Richtler, T.\ 2005, \aap, 433, 43
\reference{} Durrell, P.~R., Harris, W.~E., Geisler, D., \& Pudritz, R.~E.\ 1996a, \aj, 112, 972 
\reference{} Durrell, P.~R., McLaughlin, D.~E., Harris, W.~E., \& Hanes, D.~A.\ 1996b, \apj, 463, 543
\reference{} Evans, N.~W., Wilkinson, M.~I., Perrett, K.~M., \& Bridges, T.~J.\ 2003, \apj, 583, 752
\reference{} Fall, S.~M., \& Rees, M.~J.\ 1977, \mnras, 181, 37P
\reference{} Freedman, W.~L., et al.\ 2001, \apj, 553, 47
\reference{} Geha, M., Guhathakurta, P., \& van der Marel, R.~P.\ 2003, \aj, 126, 1794 
\reference{} Geha, M., Guhathakurta, P., \& van der Marel, R.~P.\ 2002, \aj, 124, 3073
\reference{} Gonz{\' a}lez, J.~J.\ 1993, Ph.D.~Thesis, Lick Observatory, University of California, Santa Cruz
\reference{} Graham, A.~W., Jerjen, H., \& Guzm{\' a}n, R.\ 2003, \aj, 126, 1787
\reference{} Gratton, R., Sneden, C., Carretta, E., \ 2004, \araa, 42, 385
\reference{} Gunn, J.~E., \& Gott, J.~R.~I.\ 1972, \apj, 176, 1
\reference{} Harris, G.~L.~H., Harris, W.~E., \& Geisler, D.\ 2004, \aj, 128, 723 
\reference{} Heisler, J., Tremaine, S., \& Bahcall, J.~N.\ 1985, \apj, 298, 8 
\reference{} Jerjen, H., Kalnajs, A., \& Binggeli, B.\ 2000, \aap, 358, 845 
\reference{} Jerjen, H., Binggeli, B., \& Barazza, F.~D.\ 2004, \aj, 127, 771 
\reference{} Kundu, A., \& Whitmore, B.~C.\ 1998, \aj, 116, 2841 
\reference{} Kuntschner, H., Ziegler, B.~L., Sharples, R.~M., Worthey, G., \& Fricke, K.~J.\ 2002, \aap, 395, 761
Larsen, S.~S.\ 1999, \aaps, 139, 393
\reference{} Larsen, S.~S., Brodie, J.~P., Huchra, J.~P., Forbes, D.~A., \& Grillmair, C.~J.\ 2001, \aj, 121, 
2974 
\reference{} Lee, M.~G., Kim, E., \& Geisler, D.\ 1998, \aj, 115, 947 
\reference{} Lee, H.C., Worthey, G., 2005, preprint (astro-ph/0504509) 
\reference{} Lotz, J.~M., Miller, B.~W., \& Ferguson, H.~C.\ 2004, \apj, 613, 262 
\reference{} Lotz, J.~M., Telford, R., Ferguson, H.~C., Miller, B.~W., Stiavelli, M., \& Mack, J.\ 2001, \apj, 552, 572
\reference{} Mao, S., \& Mo, H.~J.\ 1998, \mnras, 296, 847
\reference{} Maraston, C.\ 1998, \mnras, 300, 872 
\reference{} Mastropietro, C., Moore, B., Mayer, L., Debattista, V.P., Pifferetti, R., Stadel, J., 2005, preprint (astro-ph/0411648)
\reference{} Miller, B.~W., Lotz, J.~M., Ferguson, H.~C., Stiavelli, M., \& Whitmore, B.~C.\ 1998, \apjl, 508, L133  
\reference{} Moore, B., Lake, G., \& Katz, N.\ 1998, \apj, 495, 139 
\reference{} Moore, Katz \& Lake 1996
\reference{} Oke, J.~B., et al.\ 1995, \pasp, 107, 375
\reference{} Oke, J.~B.\ 1990, \aj, 99, 1621 
\reference{} Olsen, K.~A.~G., Miller, B.~W., Suntzeff, N.~B., Schommer, R.~A., \& Bright, J.\ 2004, \aj, 127, 2674 
\reference{} Pedraz, S., Gorgas, J., Cardiel, N., S{\' a}nchez-Bl{\' a}zquez, P., \& Guzm{\' a}n, R.\ 2002, \mnras, 332, L59
\reference{} Peng, E.W. et al.~2005, ApJ in press (astro-ph/0509654)
\reference{} Peng, E.~W., Ford, H.~C., \& Freeman, K.~C.\ 2004, \apj, 602, 705
\reference{} Proctor, R.~N., Forbes, D.~A., \& Beasley, M.~A.\ 2004, \mnras, 355, 1327 
\reference{} Proctor, R.~N.~\& Sansom, A.~E.\ 2002, \mnras, 333, 517 
\reference{} Puzia, T.~H., Kissler-Patig, M., Brodie, J.~P., \& Schroder, L.~L.\ 2000, \aj, 120, 777
\reference{} Rakos, K., \& Schombert, J.\ 2004, \aj, 127, 1502
\reference{} Richardson, S. \& Green, P.G., 1997, JR Statist. Soc. B, 59, 731
\reference{} Richtler, T., et al.\ 2004, \aj, 127, 2094
\reference{} Sakai, S., et al.\ 2000, \apj, 529, 698 
\reference{} Sandage, A., Binggeli, B., \& Tammann, G.~A.\ 1985, \aj, 90, 1759 
\reference{} Sargent, W.~L.~W., Schechter, P.~L., Boksenberg, A., \& Shortridge, K.\ 1977, \apj, 212, 326 
\reference{} Schiavon, R.~P., Rose, J.~A., Courteau, S., \& MacArthur, L.~A., 2005, preprint (astro-ph/0504313)
\reference{} Schlegel, D.J., Finkbeiner, D.P., \& M. Davis,  ApJ, 500, 525 
\reference{} Secker, J., \& Harris, W.~E.\ 1993, \aj, 105, 1358
\reference{} Seth, A., Olsen, K., Miller, B., Lotz, J., \& Telford, R.\ 2004, \aj, 127, 798
\reference{} Sirianni, M., Jee, M.J., Benitez, N., Blakeslee, J.P., Martel, A.R., Clampin, M., de Marchi, G., 
Ford, H.C., Gilliland, R., Hartig, G.F., Illingworth, G.D., Mack, J., McCann, W.J., \& Meurer, G. 2005, 
PASP, submitted
\reference{} Strader, J., Brodie, J.~P., \& Forbes, D.~A.\ 2004, \aj, 127, 295 
\reference{} Strader, J., Brodie, J.~P., Spitler, L., Beasley M.~A., 2005, AJ,submitted
\reference{} Thomas, D., Maraston, C., \& Korn, A.\ 2004, \mnras, 351, L19 
\reference{} Thomas, D., Maraston, C., \& Bender, R.\ 2003, \mnras, 339, 897 (TMB03) 
\reference{} Trager, S.~C., Faber, S.~M., Worthey, G., \& Gonz{\' a}lez, J.~J.\ 2000, \aj, 119, 1645 
\reference{} Tremaine, S.~D., Ostriker, J.~P., \& Spitzer, L.\ 1975, \apj, 196, 407
\reference{} Tully, R.~B., Somerville, R.~S., Trentham, N., \& Verheijen, M.~A.~W.\ 2002, \apj, 569, 573
\reference{} van den Bergh, S.\ 2000, \aj, 119, 609
\reference{} van der Marel, R.~P.\ 1994, \mnras, 270, 271 
\reference{} West, M.~J., C{\^ o}t{\' e}, P., Marzke, R.~O., \& Jord{\' a}n, A.\ 2004, \nat, 427, 31 
\reference{} White, S.~D.~M., \& Frenk, C.~S.\ 1991, \apj, 379, 52
\reference{} Worthey, G.~\& Ottaviani, D.~L.\ 1997, \apjs, 111, 377
\reference{} Worthey, G., Faber, S.~M., Gonzalez, J.~J., \& Burstein, D.\ 1994, \apjs, 94, 687
\reference{} van Zee, L., Barton, E.~J., \& Skillman, E.~D.\ 2004a, \aj, 128, 2797  
\reference{} van Zee, L., Skillman,E.~D., \& Haynes, M.~P.\ 2004b, \aj, 128, 121 
\reference{} Vesperini, E.\ 1997, \mnras, 287, 915 
\reference{} Zepf, S.~E., Beasley, M.~A., Bridges, T.~J., Hanes, D.~A., Sharples, R.~M., Ashman, K.~M., \& Geisler, D.\ 2000, \aj, 120, 2928 
\end{references}
\end{document}